# The Perils of Kremlin's Influence: Evidence from Ukraine[*]

Chiara N. Focacci      Mitja Kovac      Rok Spruk

## Abstract

*We examine the contribution of institutional integration to the institutional quality. To this end, we exploit the 2007 political crisis in Ukraine and examine the effects of staying out of the European Union for 28 Ukrainian provinces in the period 1996-2020. We construct novel subnational estimates of institutional quality for Ukraine and central and eastern European countries based on the latent residual component extraction of institutional quality from the existing governance indicators by making use of Bayesian posterior analysis under non-informative objective prior function. By comparing the residualized institutional quality trajectories of Ukrainian provinces with their central and eastern European peers that were admitted to the European Union in 2004 and after, we assess the institutional quality cost of being under Kremlin's political influence and interference. Based on the large-scale synthetic control analysis, we find evidence of large-scale negative institutional quality effects of staying out of the European Union such as heightened political instability and rampant deterioration of the rule of law and control of corruption. Statistical significance of the estimated effects is evaluated across a comprehensive placebo simulation with more than 34 billion placebo averages for each institutional quality outcome.*

**Introduction**

The notion that institutional integration matters for economic growth and development has become well-established in the scholarly literature and policy debate (Rivera-Batiz and Romer 1991, Eichengreen 2007, Campos et. al. 2019). Scholars traditionally distinguish between two types of integration (Lawrence 1996). A shallow form of integration is usually based on trade liberalization without affecting regulatory and structural policies. By contrast, a deep institutional integration is connoted not only with trade liberalization but also with broader policies such as competition and regulation and entails deeper provisions. The economic effects of institutional integration are relatively well-understood. For instance, Campos et. al. (2019) examine the economic growth effects of the European Union membership, find positive but heterogenous effects in a sample of non-founding member states, and show that trade openness, financial development and the adoption of single currency accounts for these effects. Similar positive but heterogeneous effects of the membership in the European Union have been confirmed by Maudos et. al. (1999), Badinger (2005), Crespo Cuaresma et. al. (2008) and Garoupa and Spruk (2020) among several others.

The *raison d'être* of these studies is the search for the counterfactual scenario of the institutional integration. By making use of the methods such as difference-in-differences or synthetic control estimator (Abadie et. al. 2015, Maseland and Spruk 2022), an appropriate policy counterfactual scenario can be estimated either in the presence or absence of parallel trends. The general thrust of such approach is the comparison of countries that have entered the institutional integration to the appropriate donor pool of countries that have not been admitted therein. By reproducing the pre-entry trajectory of economic outcomes, post-entry difference between the observed trajectory and its hypothetical counterpart can be observed to assess the impact of the institutional integration.

Several limitations arise from the existing approach to estimate the effects of institutional integration. First, whilst the search for counterfactual scenario and its estimation may uncover the effects of admission to the institutional integration for the countries that joined, the question that remains less clear is simple. Namely, what is the effect of staying out of the institutional integration? If a certain type of institutional integration promulgates growth, trade and specialization, estimating the costs of staying out of such growth-enhancing type of integration would be more informative than estimating the benefits of admission. And second, most scholarly studies deal with the economic effects of institutional integration. Perhaps the most obvious question to ask beforehand is whether the inclusion in the institutional integration either improves or deteriorates the institutional quality such as the rule of law, efficacy of public administration and the control of corruption to better understand whether the institutional integration is a curse or cure for the institutional quality. In this paper we fill these two voids in the literature.

In particular, we examine the effects of staying out of the European Union on the institutional quality for Ukrainian provinces between the years 1996 and 2020. We first estimate the institutional quality of Ukrainian provinces by estimating the series of governance indicators through the extraction of residual component from the existing aggregate governance series. In the next step, we apply Bayesian posterior analysis through the use of Metropolis-Hastings random-walk algorithm to compute the ideal points of institutional quality, and smooth the series through the high-frequency filtering-based variance decomposition. By comparing the full set of Ukrainian provinces to the full sample of regions from Central and Eastern Europe that have been admitted to the European Union until 2020, we are able to assess the institutional quality effects of staying out of the European Union which may inform us about the potential institutional quality loses of Kremlin's political influence as a proxy for shallow institutional integration. Our contribution is of reasonable relevance for policymakers to better understand what actions to take in view of the EU membership potentially being granted to Ukraine. In particular, it sheds clarity on the disastrous economic consequences that await Ukraine in case of Russia coming to power and expectantly contributes to reaching sound policy making decisions for both Ukraine and Europe as a whole. The article also contributes to the extant literature on the effect brought by Russian institutions abroad.

On 24 February 2022, Russia started its military invasion of Ukraine, which progressively escalated in the first war Europe experienced in decades. While it is not possible to estimate casualties among civilians –although, the Emergency Service counts more than 4,000–, the United Nations High Commissioner for Refugees claims more than 14 million Ukrainian people have already fled the country, giving origin to one of the largest refugee crises since World War Two. In parallel, European states have imposed tough financial sanctions on Russia. However, penalties brought on Russia and the military Ukrainian resistance may not be sufficient to avoid Russia taking full or partial control of Ukraine. While political experts discuss the possibility of a larger conflict, Ukraine's Prime Minister Zelenskyy ups pressure on the European Union for immediate membership in a desperate attempt to avoid Russia taking over.

In his seminal work, Lynch (2001) illustrated how, in general, Russian influence is usually more or less successful where the Western stake is least or most intense. In an attempt to 'maintain the appearance of great power status abroad' after the 1991 decay of the Soviet Union, Russia has been obliged to remain active in international commitments, including political consultations with respect to both the Balkans and NATO. On this subject, Tolstrup (2009) observed that in the post-Soviet area, the foreign policy by Russia in the 'Near Abroad' –including Belarus, Armenia, Azerbaijan, the Central Asian republics, the Baltic States, Moldova, Ukraine, and Georgia– has had a significantly negative impact on democratic perspectives, including the civil and political rights of people. Furthermore, following the Kremlin's annexation of Crimea, Götz (2016) provided evidence for Russia's 'increasingly assertive behavior in the post-Soviet space'.

After the disintegration of the Soviet Union, Russia's attempts to re-establish its identity abroad have been more and more frequent. By counting on supporting Russian diaspora populations, Russia has exploited public diplomacy to persuade foreign publics top-down (Just 2016). As a result, Putin managed to reinforce 'the populist desire to see Russia return to the world stage as an economic and military superpower' (Mersol 2017), particularly so where hybrid or authoritarian polities persist in the non-Baltic post-Soviet space (Cameron and Orenstein 2012). Contrary to China, Russia seems to encourage an ethnocentric viewpoint of a country and a regime that need to be defended against Western threats (Popovic et al. 2020). With the recent invasion of Ukraine, Russia's international reputation has certainly worsened and, at the same time, confirmed how their actions directly derive from an evident fear of Western influence in the Near Abroad (Kerrane 2020).

In 1991, Ukraine formally declared its independence from Soviet Union. Early institutional and political development after the demise of the Soviet Union was characterized by the conflicting policy overtures to the Western nations and alliance with Russia leading to the political crisis. In 2004, the opposition leader Viktor Yushchenko launched a mass protest campaign in the light of rigged elections giving victory to pro-Russian candidate Viktor Yanukovich. The election results were subsequently nullified by the Supreme Court of Ukraine. By the end of 2005, Viktor Yushchenko won the elections leading to the souring relations with Russia and frequent gas supply and pipeline shortages and disruption. In 2006 parliamentary elections, power struggles and tensions evolved between the pro-Western president and the pro-Russian parliamentary majority leading to the dissolution of the parliament. Early in 2007, several members of the opposition launched the support to the ruling coalition leading to a high probability of securing two-thirds majority that would empower the parliament to override the president's veto right and, thus, initiate constitutional changes. In 2007, the president decreed the dissolution of unicameral parliament known as *Verkhovna Rada*. President's authority to dismiss the parliament was immediately challenged in the Supreme Court whilst the president intervened directly into the operation of the Constitutional Court through the dismissal of two judges ensuring that the court has not ruled on the constitutionality of the decree. In the elections, the coalition of two electoral blocs associating themselves with the Orange Revolution won a narrow majority. Yet, significant regional variation occurred between the western Ukraine, predominantly voting in favour of the pro-Western electoral bloc and eastern Ukraine, where the pro-Russian coalition obtained the majority of popular vote amid the allegations of electoral fraud. Although the elections generally met international standards, observers noted delays in the formation of district election commissions, inadequate quality of voter lists, disenfranchisement of voters due to law amends that abolished absentee ballots, removed voters from the ballot list that have crossed the border after August 1, 2007, modalities for voting at home, and extensive campaigning by state and local officials in violation of the law.

The uniqueness of the 2007 political crisis as prelude to the subsequent institutional instability can be hardly neglected. In Figure 1, we formally test whether the 2007 political crisis posits a structural break. From Worldwide Governance Indicators databse (Kaufmann et.al. 2011), we compute the first principal component of the institutional quality from six baseline indicators,[2] and apply a simple version of the synthetic control estimator (Abadie et. al. 2015) to compare Ukraine's trajectory of latent institutional quality with a donor pool of 101 countries for the period 1996-2016. The evidence indicates a pervasive structural break in response to 2007 political crisis, permanently derailing the observed institutional quality of Ukraine producing a large-scale structural breakdown. That said, the observed institutional quality of Ukraine has worsened considerably after the 2007 political crisis compared to its control group[3] that contains similar pre-2007 attributes. The average institutional quality effect of the 2007 political crisis is around -0.98 which entails a large-scale downward deviation from pre-2007 equilibrium. In this paper, we exploit the 2007 political crisis as a source of counterfactual scenario and estimate the institutional quality of Ukraine and its region in the hypothetical case of the admission to the European Union in temporally similar round of enlargement to the rest of Central and Eastern European countries. Since Ukraine remained outside the European institutional integration after 2007 both on de jure and de facto terms, the resulting gap in institutional quality can partially reflect the institutional quality effect of staying out of the European Union and being prone to the policy overtures under Kremlin's influence and subsequent political and institutional destabilization attempts. Our results uncover massive institutional quality benefits of the European integration. Our estimates indicate substantial improvements of political stability, control of corruption, liberal democratic governance and rule of law that is entirely comparable with post-enlargement institutional quality development in the central and eastern European peers. Our identification strategy allows to almost entirely remove the pre-treatment imbalance in outcomes and covariates, thus giving credence to the post-2007 effects reflecting the institutional quality cost of staying out of the European Union rather than the misfit, imbalance or alternative institutional and policy shocks.

---

[2] Voice and accountability, political stability and absence of violence, government effectiveness, regulatory quality, rule of law, and the control of corruption

[3] The institutional quality trajectory of Ukraine before 2007 political crisis is best reproduced as convex combination of institutional quality and the auxiliary geographic, economic and demographic characteristics of Serbia (31 percent), Paraguay (19 percent), Argentina (16 percent), Romania (15 percent), Indonesia (13 percent) and Iraq (6 percent), respectively. The predictive discrepancy between Ukraine and its synthetic control group is low with root mean square prediction error (RMSE) of around 0.051.

FIGURE 1 – INSERT ABOUT HERE

In a 2014 meeting of the Royal Institute of International Affairs, Chatham, experts have included the desire of the Kremlin to expand the Russian World and solidify a Russian style of governance in the countries 'in its sphere of influence' as the ultimate goal of Russian foreign policy. Today, Ukraine serves as an example of this goal, risking an economic –and political– crisis similar to that experienced by other states before her. Projections indicate that Europe's GDP is likely to fall by 4 percentage points, also due to the cut of Russian gas supply, while inflation is expected to rise by 1.5 percentage points. With respect to what awaits Ukraine more specifically, some evidence is provided by the extant literature on other cases that escaped or surrendered to Russian domination.

In terms of contrasting example, the Polish economy, for instance, significantly grew in the post-Soviet period (Slay 2014), especially thanks to boosts in foreign trade following EU membership (Kowalski and Shachmurove 2018), and also at the local level (Young and Kaczmarek 2000). Lithuania, Estonia, and Latvia managed to successfully turn to capitalism while industrial relations in Belarus did not progress substantially due to the system of ruling elites and their attitude towards Russia (Savchenko 2002). The Commonwealth of Independent States (Russia, Belarus, and Ukraine) is, in general, also less integrated with the world economy, having led to a decline in the manufacturing and agricultural sectors (Lane 2011). When looking at trade matters, Isakova et al. (2016) provided evidence that the constitution of a customs union between Belarus, Kazakhstan, and Russia has only and partially benefitted Russia, while it has negatively impacted imports from China and the EU to, respectively, Kazakhstan and Belarus. For the latter, this is also the result of a political economy that started disincentivizing entrepreneurial social groups (Yarashevich 2014). With our analysis, we aim to provide some answers to a widely debated issue; namely, whether a similar destiny awaits Ukraine if Russia were to take over.

The paper is structured as follows. Section 2 provides the historical and institutional background discussion. Section 3 presents the data and variables used in our analysis. Section 4 discusses the identification strategy. Section 5 discusses the results and robustness checks. Section 6 concludes.

## 2  Background

As well-illustrated by Wydra (2004), Ukraine has tried to find independence from the larger and more powerful neighbour Russian state since August 1991, when the first tensions started to emerge. Months after declaring the independence of Republic of Crimea, the December referendum showed only 54% of the voters in Crimea in favour of Ukraine's independence. In parallel, Russian populist leader Zhirinovsky and others claimed ownership of Crimea, considered a strategic port for the Black Sea Fleet. Tensions increased in 1994, when the office of President of Crimea was set up in the person of Meshkov, who pushed for a Crimean Constitution and rights autonomous from Ukraine. A Constitution accepted by the Ukrainian government would only be put in place in 1998, when it was guaranteed that relations with the Russian communist party would be suspended and that local decisions would not contradict Ukrainian laws. This, however, did not exclude the predominance of the Russian language and influence in the Crimean population.

Neither Ukraine nor Russia initially recognized reciprocal independence, but rather insisted on each other's 'sovereignty', with Ukraine giving in to an economic union with Russia to avoid the economic pressures of isolation. At the time, the Minister Kuchma saw access to the European Community unlikely (Morrison 1993). In 1997, the bilateral inter-state treaty between Ukraine and Russia was signed in favour of the inviolability of the countries' existing borders coming in effect in 1999, with member of the Russian parliament and communist party Seleznev calling it 'a great victory for Russia' (Tolz 2002).

The revisionist voice of Russian politics would make it progressively more difficult to strengthen the ties with the West. Following the explosion of the pro-Western Orange revolution and consequent elections of President Yushchenko in 2005, the imperialistic powers of the Kremlin made it clear in 2009, when the Duma passed legislation 'to permit Russian forces to intervene abroad in defence of Russian citizens', that Ukraine would represent 'a real driving force' towards the West and, therefore, a threat to the Soviet power (Larrabee 2010). In this period –when gas tensions between Ukraine and Russia also become more frequent– it became progressively clear that the idea of a Russian world would be positively and significantly accepted only in areas of predominantly ethnic Russians or Russian speakers, who perceived the Kremlin's assertive policy as protective of their community (Feklyunina 2016).

Further tensions emerged in April 2014, when Russia withdrew the gas price discount granted to Ukraine under President Yanukovich, elected in 2010, and claimed for the repayment of Ukraine's debt (Loskot-Strachota and Zachmann 2014). This followed the coup of February against President Yanukovich, committed to a national pragmatism, or the establishment of good or neutral relations with Russia, for the sake of Ukraine's future in Europe (Kolosov 2018). In June, the Association Agreement between the European Union and Ukraine was finalised, with President Poroshenko stating that being in the EU 'would mean the world to his country' (Pridham 2014). Overall, following the rather optimistic and cooperative 1992-94 period, the time of troubles of 1994-2000, when the Agreement on Partnership and Cooperation was signed, insisting on 'democracy and human rights as an essential element of the PCA', and the Putin era from 2000 onwards, 2014 marked a year of non-return for Ukraine (and EU) relations with Russia.

Unable to create a stable and cooperative regime in Kyiv, Russia progressively opted for a solidification of local regimes' dependence on Russian support so as to maintain control over parts of the country, including efforts at state-buildings in Donbas (Malyarenko and Wolff 2018). This culminated in the illegal annexation of the Republic of Crimea and Sevastopol to Russia in March 2014, with a referendum that declared 96.7% of Crimean voters in favour. The growing fighting in the Donbas area pushes Ukraine's interest in joining the EU and NATO to an even larger extent. Following several cyberattacks in Ukraine on behalf of Russia, current President Zelenskyy is elected in a context of a stagnating economy and an ongoing conflict with Russia, worsened by US President Trump's suspension of military aid.

After Russian Armed Forces started massing in Crimea and at the border with Ukraine in the spring of 2021, the level of tension between Ukraine and Russia became alarming when Putin recognised the Donetsk People's Republic and the Luhansk People's Republic in February 2022. Thousands of thousands of troops were sent to the borders of Ukraine on the morning of 24 February, explicitly demanding European countries and the US to bar Ukraine from ever joining NATO. The request was not accepted as the devastating military assault from territories of Russia, Belarus, and Crimea in Ukraine continued to destroy cities, kill civilians (for a total of 4,169 as of June 1, 2022), and produce mass migration to neighbouring countries. This ultimately forced the US, UK, and EU member states to join forces and implement sanctions targeting Russian oligarchs and banks. While the consequences of a Russian take over in Ukraine might be evident, it remains crucial to provide empirical evidence on the perils of the Kremlin's influence on today's Ukraine and its future.

## 3    Data

### 3.1    Measuring institutional quality

Our approach to measure institutional quality in Ukrainian regions is constrained by the lack of observable and measurable characteristics to provide insights into the variation in institutional quality. To address these concerns, our empirical strategy to measure subnational institutional quality relies on the extraction of the residual component of the institutional quality from the observable aggregate institutional quality series using the variation in pre-determined geographic characteristics. Therefore, our aim is to extract a latent variable component from higher-level aggregation and project it to the subnational-level using the set of pre-determined characteristics that cannot be empirically manipulated.

Our approach is similar to Magnusson and Tarverdi (2020) method of estimating governance quality which we combine with the plausibly orthogonal characteristics used for a linear projection of institutional quality to the subnational level. Our set of aggregate institutional quality variables consists of well-established Worldwide Governance Indicators (Kaufmann et. al. 2011) and encompass five distinctive variables. First, voice and accountability component captures the perception of the extent to which the citizens are able to participate in the selection of government and the extent to which they enjoy freedom of expression, association and media. This particular dimension partially captures the degree of liberal democracy that can be compared across space and time. Second, political stability and absence of violence component indicates the perception of the likelihood that the government will be destabilized or overthrown by unconstitutional or violent means that includes politically-motivated violence and terrorism. This particular dimension largely captures the degree of political and institutional stability.

Third, government effectiveness component captures the overall quality of public services and civil service, and the associated degree of independence from political pressures, the quality of policy formulation, and implementation and government's credibility and commitment to such policies. This particular dimension largely captures the institutional effectiveness and quality of bureaucracy, policy stability, quality of government administration and public goods provision. Fourth, regulatory quality component indicates the ability of government to formulate and implement sound economic and structural policies that facilitate and enhance private sector development. It encompasses the presence of policy distortions such as unfair competitive practices, price controls, discriminatory tariffs, excessive trade and non-trade protection and barriers, discriminatory taxes, state ownership and intervention in the private sector, extent of labor regulation as well as the efficacy and enforcement of competition regulation. This particular dimension largely captures the institutional and policy constraints on private sector development. Fifth, rule of law component captures the extent to which citizens have confidence in and abide by the rules of society with respect to contract enforcement, police, courts and the likelihood of crime and violence. The rule of law dimension encompasses the fairness and speediness of the judicial process, private property protection, judicial independence, efficiency of legal framework, intellectual property rights protection as well as trust in courts and police. It captures the strength of the rule of law and public trust in institutions. And fifth, control of corruption captures the extent to which public power is exercised for private gains through petty and grand forms of corruption as well as state capture by economic and political elites and private interests. Control of corruption measures is constructed from the variables indicating corruption among public officials, diversion of public funds, transparency, accountability and corruption in the public sector, strength and enforcement of anti-corruption policy, and the frequency of paying bribes to various branches of public sector.

Henceforth, we exploit the variation in each dimension of institutional quality using the updated version of Kaufmann et. al. (2011) *Worldwide Governance Indicators* data. Following Magnusson and Tarverdi (2020), our goal is to extract the residual component of each governance indicator from the aggregate level based on the variation. From the aggregate governance score, we use the set of pre-determined time-invariant covariates that are orthogonal to the institutional quality of interest. This approach allows us to residual the aggregate institutional quality score to the regional level. Without the loss of generality, suppose we observe a finite sequence of countries indexed by $i = 1,2,...N$ that is observed over time $t = 1,2,...T$ and a continuum of regions $j = 1,2,...J$ where $i \in N \in J$. For each $t \in T$ and the full cross section of $j = 1,2,...J$ regions, we estimate a simple canonical regression of the following form:

$$q_{j \in J} = \theta_0 + \mathbf{X}'_{j,i \in N}\beta + \varepsilon_{j,i \in N} \qquad (1)$$

where $q$ denotes the institutional quality variable of interest, and $\mathbf{X}$ represents the vector of pre-determined physical geographic covariates such as latitude, longitude, landlocked indicator, precipitation, mean temperature and elevation level as well as the indicators of climatic zone based on the Köppen-Geiger climate classification scheme (Kottek et. al. 2006, Peel et. al. 2007). The random error is denoted by $\varepsilon$ and captures the stochastic disturbances and transitory shocks to institutional quality with i.i.d. structure. From the canonical regression in Eq. (1), we compute the residual component of the respective dimension of institutional quality for the full section of Ukrainian regions and its donor pool, both of which are indexed by $j = 1,2,...J$:

$$e_{i \in N} = \tilde{q}_{i \in N,j} - \bar{q}_{i \in N,j} \qquad (2)$$

where $\tilde{q}$ denotes the observed realization of institutional quality and $\bar{q}$ represents the predicted level of institutional quality. The recovered residual component provides an important guide behind the interpretation of the series. First, a positive residual component invariably suggests that the level of institutional quality observed in *j*-th treated and control region is higher than the level plausibly expected in geographically similar areas at the regional level. Secondly, a zero residual component indicates the level of institutional quality at the same level as expected in geographically similar environment. And thirdly, a negative residual component suggests that the implicit level of institutional quality is worse than expected in places with geographically similar conditions. Such spatial variation in the residual component is consistent with the notion of institutional quality delineated between the inclusive-participatory type and the extractive-exclusionary one popularized by Acemoglu et. al. (2005). Hence, the residual component can be used to determine the spatial disparities in institutional quality based on the difference between the observed and predicted level of institutional quality. The substantive insight of the residual component may indicate if the observed institutional quality appears to be more inclusive and participatory than expected in geographically similar endowments, or if it appears to entail a cluster of more exclusionary economic and political institutions. The overall level of institutional quality is recovered from the observed aggregate component across $i = 1,2,\ldots N$ dimension and the exogenous component that captures region-level idiosyncratic institutional specificities:

$$Q_{i \in N} = q_{j \in J} + e_{i \in N} \tag{3}$$

where $Q$ denotes the overall institutional quality, $q$ is the observed institutional quality in j-th region projected from $i = 1,2,\ldots N$ level of aggregation, and $e$ is the exogenous residual component. Our primary interest lies in the spatial and temporal variation in $e$ which reflects region-level institutional idiosyncrasies and denotes whether the observed quality is better or worse than the expected one based on the variation in pre-determined physical geographic characteristics.

One of the key limitations behind the repeated cross-sectional estimation of the residual component scores of institutional quality concerns the time invariance of geographic characteristics such as latitude and longitude or binary indicators of coastal access and landlocked status. This implies that, the resulting variation in the residual component may only shift the intercept $\theta_0$ upward or downward whilst rendering it constant over time. Under these circumstances, most of the variation in *e* would be absorbed by the unobserved effects, precluding any panel-level analysis.

Albeit imperfectly, we partially overcome these limitations by making use of Monte Carlo Markov Chain (MCMC) sampling algorithm for the probability distribution of the recovered residual component of institutional quality. By constructing the Markov chain with the specified distribution as the equilibrium, a sequenced sample of the target distribution can be obtained by recording the states of institutional quality from the chain (Jarner and Roberts 2007). Our aim is to obtain an uninterrupted sequence of random samples from the pre-specified probability distribution where random sampling is difficult since our data is observational per se. We may build such sequence of random samples by adopting the well-known Metropolis et. al. (1953) and Hastings (1970) algorithm approximating the true distribution of institutional quality at the regional level indexed by $j = 1,2,..J$. Notice that randomly sequenced samples can be drawn from any probability distribution that has a known probability density function that is proportional to the density to the target density function. The general intuition behind the algorithm is that the generated random sequence of samples can be generated in such a way that as more samples are produced, the observed distribution of institutional quality parameter more closely approximates the target distribution through an iterative procedure that creates the Markov Chain. At each respect iteration, the algorithm selects a candidate for the proximate sample value from the current sample. If the probability threshold is specified ex-ante, the candidate value of institutional quality is either accepted or rejected if it exceeds the threshold.

Let $f(Q = e_{i \epsilon N})$ be a density function proportion to the target probability distribution denoted by $P(Q = e_{i \epsilon N})$. Our goal is to obtain a posterior parameter of residualized institutional quality at the regional level to be evaluated, denoted by $\pi(\Theta_i) = p(\Theta_i | \Theta_{i-1}, X, Q)$. We assume a single parameter to be sampled from a one-dimensional distribution space characterised by $\pi(\Theta)$. To generate a sequence of random samples from $\pi(\Theta)$, the algorithm must specify a plausible probability density function denoted as $q(\Theta^{g+1} | \Theta^g)$ with the implicit required density ratio, $\frac{\pi(\Theta^{g+1})}{\pi(\Theta^g)}$. In this respect, the algorithm draws a candidate value first that will be either accepted or rejected based on the acceptance probability. Similar to the Gibbs sampling algorithm, our approach consists of two-stage sampling procedure:

**Step #1**: Draw $\Theta^{g+1}$ from the proposal density $q(\Theta^{g+1}|\Theta^g)$

**Step #2**: Accept $\Theta^{g+1}$ with probability $\gamma(\Theta^{g+1}|\Theta^g)$ where: $\gamma(\Theta^{g+1}|\Theta^g) = \min\left(\frac{\pi(\Theta^{g+1})}{q(\Theta^{g+1}|\Theta^g)} \bigg/ \frac{\pi(\Theta^g)}{q(\Theta^g|\Theta^{g+1})}, 1\right)$

where the key requirement to generate a sequence of random samples is that the posterior parameter values are drawn from the proposed distribution of the random variable to evaluate the acceptance and rejection criteria. Using two designated steps, the algorithm decomposes the unrecognizable conditional distribution into a recognizable one through the randomly sequence generation of candidate points, and the unrecognizable part from which the acceptance criteria is set. Such iterative procedure adjusts the equilibrium distribution by specifically allowing for a non-analytic functional form of the model. One particular advantage of the algorithm is that sampling can be conducted on a specific tail of the distribution to analyse parameter restrictions imposed by the prior values. However, imposing prior values to restrict parameters in space requires a nearly complete information matrix and may also be driven by subjective beliefs. Henceforth, our approach is to abstain from the imposition of priors. To facilitate the generation of random samples through the Markov chain, our approach is to impose a non-informative objective prior distribution on parameter space (Jeffreys 1946) where the density function is proportional to the square root of the determinant from Fisher information matrix $p(\vec{\Theta}) \propto \sqrt{\det \mathcal{J}(\vec{\Theta})}$. A notable advantage of the non-informative prior is posited by the non-thin tails of the distribution compared to its target counterpart, ensuring a straightforward and fast convergence of the algorithm.

By imposing objective non-informative prior to approximate the shape of the target distribution of institutional quality, we adopt an adaptive random-walk version of the Metropolis-Hastings algorithm where a candidate value of institutional quality residual for each region is drawn from a simple random walk model characterized by $\Theta^{g+1}_{i\epsilon N,t} = \Theta^{g}_{i\epsilon N,t} + \varepsilon_{i\epsilon N,t}$ where $\varepsilon_{i\epsilon N,t} = i.i.d \sim (0, \sigma^2)$ is the random component assumed to be a symmetric density function with thick tails of t-distribution. Our choice of the density function is generic. Therefore, it ignores the structural features of the target density. Given the symmetry in the proposed density functions, $q(\Theta^{g+1}|\Theta^g) = q(\Theta^g|\Theta^{g+1})$, the algorithm used to estimate the subnational institutional quality converges to the following two-stage procedure:

**Step #1**: Draw $\Theta^{g+1}$ from the proposal density $q(\Theta^{g+1}|\Theta^g)$

**Step #2**: Accept $\Theta^{g+1}$ with probability $\gamma(\Theta^{g+1}, \Theta^g)$ where: $\gamma(\Theta^{g+1}, \Theta^g) = \min[\pi(\Theta^{g+1})/\pi(\Theta^g) t, 1]$

Two chief advantages arise in estimating the latent institutional quality. First, the simplified algorithm has the ability to control the variance of the error term which further reduces the measurement error. And second, the algorithm must be adjustable for the variance of the error term to obtain an acceptable level of accepted draws. Given a relatively large size of our sample, we set the number of iterations to obtain a sequence of random samples at 12,500 for each outcome-year combination and set the acceptance rate at 25 percent, which appears to be in the conventional range between 20 percent and 40 percent. At 12,5000 iterations per outcome and year, the number of discarded observations through the burn-in amounts to 2,500. This implies that for 24 years and six outcomes under consideration, the overall number of iterations performed in our analysis is around 1,800,000 million randomly sequenced samples. Therefore, for each $t = 1,2,...T$, we construct the mean and median MCMC estimate of the posterior parameter. By inverting the test statistics, the upper and lower confidence bounds are construct for the full panel.

One of the intrinsic limitations behind the random-walk version of Metropolis-Hastings algorithm arises from the potential instability and composition of the latent component over time due to the simultaneous presence of noise and signal in the series. Estimating a chained sequence of randomly generate samples through MCMC algorithm typically invokes a confluence of both cyclical and deterministic component that may not disappear when the non-informative objective prior function is introduced. Whilst the deterministic component captures long-run behaviour and trend of the institutional quality series, the cyclical component may capture short-run transitory deviations of the posterior parameter values from its long-run equilibrium. For the sake of brevity, suppose that the posterior parameter of institutional quality for each $j = 1,2,..J$ and $t = 1,2,...T$ can be decomposed into the deterministic and cyclical component:

$$\Theta_{i,j,t} = \tau_{i,j,t} + c_{i,j,t} \qquad (4)$$

where $\tau$ denotes the deterministic (i.e. long-run) component and $c$ captures the cyclical (i.e. short-run) component of the institutional quality parameter. Without the loss of generality, it may be noted that the deterministic component captures the signal of the institutional quality parameter whereas the cyclical component reflects the noise behind the latent institutional quality trait. If the latent trait denotes transitory deviations from the long-run equilibrium, $\theta$ has a tendency to exhibit a stationary mean-reverting pattern since, it is dominated by $c$ instead of $\tau$. Therefore, if $c$ component prevails over time, the MCMC simulated latent quality trait will converge to the random-walk behaviour. By contrast, if the latent trait exhibits little deviation from the cyclical component, the underlying posterior parameter $\Theta$ is dominated by $\tau$. Our goal is modest and sets to recover a smooth long-run trend component of the posterior parameter for $j$-th region through the Hodrick and Prescott (1997) solution to the following optimization problem:

$$\min \sum_{t-1}^{T}(\Theta_{i,j,t} - \tau_{i,j,t})^2 + \phi\left((\tau_{i,j,t+1} - \tau_{i,j,t}) - (\tau_{i,j,t} - \tau_{i,j,t-1})\right)^2 \qquad (5)$$

where $(\Theta_{i,j,t} - \tau_{i,j,t})$ represents the residual component of the latent quality trait that satisfies the inner stability condition such that $c_t \sim i.i.d \sim \mathbb{N}(0,1)$. The parameter $\phi$ captures the speed of dynamic quality parameter adjustment over time to extract the cyclical component from its long-run counterpart. To avoid compressing the temporal variation, we adopt the criteria advocated by Ravn and Uhlig (2002) stating that $\phi$ should vary by the fourth power of frequency observation ratio. For annual observations, this implies that $\phi = 6.25$ ($=1600/4^4$) to smooth the series of latent institutional quality trait for considered dimension.

### 3.2 Sample

Our treatment sample consists of 24 Ukrainian provinces (i.e. oblasts)[4] and two cities with special status.[5] Our control sample comprises 195 regions from the Central and Eastern European countries that were admitted to the European Union in 2004 and later. For the 2004 enlargement round this includes Czech Republic[6], Hungary[7], Latvia[8], Lithuania[9], Poland[10], Slovakia[11] and Slovenia[12] whilst Estonia is not included.[13] For 2007 enlargement round, the control sample comprises regions from Bulgaria[14] and Romania[15]. In addition, the regions from Croatia[16] that was admitted to the European Union in 2013 are also included into the control sample. Our period of investigation comprises the years between 1996 and 2020. Our overall sample covers 5,525 region-year observations pooled into a strongly balanced panel. The size of the treatment

---

[4] Cherkasy, Chernihiv, Chernivtsi, Dnipro, Donetsk, Ivano-Frankivsk, Kharkiv, Kherson, Khmelnytskyi, Kyiv, Kirovohrad, Luhansk, Lviv, Mykolaiv, Odessa, Poltava, Rivne, Sumy, Ternopil, Vinnytsia, Volyn, Zakarpattia, Zaporizhzhia, Zhytomyr

[5] Kyiv City and Sevastopol-Crimea

[6] 14 regions: Jihocecky kraj, Jihomoravski kraj, Karlovarsky kraj, Kraj Vysocina, Karlovehradecky kraj, Liberecky kraj, Moravskoslezsky kraj, Olomoucky kraj, Pardubicky kraj, Plzensky kraj, Praha, Stredocesky kraj, Ustecky kraj, Zlinsky kraj

[7] 20 regions: Baranya, Borsod-A-Z, Budapest, Bács-Kiskun, Békés, Csongrad, Fejér, Gyor-M-S, Hajdú-Bihar, Heves, Jász-N-Sz, Komárom-E, Nógrád, Pest, Somogy, Szabolcs-Sz-B, Tolna, Vas, Veszprém, Zala

[8] 26 rajons: Aizkraukle, Aluksne, Balvi, Bauskas, Ccsu, Daugavpils, Dobeles, Gulbenes, Jckabpils, Jelgavas, Kraslavas, Kuldigas, Liepajas, Limbadu, Ludzas, Madonas, Ogres, Preiiu, Pczeknes, Riga, Saldus, Talsu, Tukums, Valkas, Valmieras, Ventspils

[9] 10 apskritis: Alytuas, Kaunas, Klapedos, Marijampoles, Panevezio, Siauliu, Taurages, Telsiu, Utenos, Vilniaus

[10] 15 regions: Dolnoslaskie, Kujawsko-Pomorskie, Lubelskie, Lubuskie, Lódzskie, Malopolskie, Mazowieckie, Opolskie, Podkarpacike, Pomorsko-Zachodnipomorskie, Slaskie, Swietokrzyskie, WarminskoMazurskie, Wielkopolskie

[11] 8 regions: Banska Bystrica, Bratislava, Kosice, Nitra, Presov, Trencian, Trnava, Zilina

[12] 12 statistical regions: Gorizia, Inner Carniola-Carso, Littoral-Carso, Carinthia, Central Slovenia, Mura, Upper Sava, Upper Carniola, Drava, Savinja, Lower Sava, Lower Carniola

[13] One of the reason behind the exclusion of Estonia is that its institutional quality scores from 1996 onwards are disproportionately high compared to its Central and Eastern European peers which implies that the values of Estonian regions may not fall within the convex hull of Ukrainian provinces institutional quality.

[14] 27 regions: Blagoevgrad, Burgas, Dobrich, Gabrovo, Haskovo, Karzhali, Kyustendil, Lovech, Montana, Pazardzhik, Pernik, Pleven, Plovdiv, Razgrad, Shumen, Silistra, Sliven, Smolyan, Sofia, Sofia Stolitsa, Stara Zagora, Targovishte, Varna, Veliko Tarnovo, Vidin, Vratsa, Yambol

[15] 42 regions: Alba, Arad, Arges, Bacau, Bihor, Bistrita-Nasaud, Botosani, Braila, Brasov, Bucharest, Buzau, Calarasi, Caras-Severin, Cluj, Constanta, Covasna, Dambovita, Dolj, Galati, Giurgiu, Gorj, Harghita, Hunedoara, Ialomita, Iasi, Ilfov, Maramures, Mehedinti, Mures, Neamt, Olt, Prahova, Salaj, Satu Mare, Sibiu, Suceava, Teleorman, Timis, Tulcea, Valcea, Vaslui, Vrancea

[16] 21 counties: Bjelovar-Bilogora, Brod-Posavina, City of Zagreb, Dubrovnik-Neretva, Istria, Karlovac, Koprivnica-Križevci, Krapina-Zagorje, Lika-Senj, Međimurje, Osijek-Baranja, Požega-Slavonia, Primorje-Gorski Kotar, Sisak-Moslavina, Split-Dalmatia, Varaždin, Virovitica-Podravina, Vukovar-Syrmia, Zadar, Zagreb County, Šibenik-Knin

sample comprises 650 region-year paired observations whilst the size of control sample comprises 4,875 paired observations.

Figure 2 depicts the evolution of Metropolis-Hastings estimated institutional quality residuals across Ukrainian provinces for the full period of our investigation. The figure exhibits mean MCMC estimates of the particular latent institutional quality trait for the Ukrainian provinces from 1996 onwards along with the 95 percent confidence bounds. Notice that the zero threshold indicates the boundary between the inclusive-participatory and extractive-exclusionary institutions characterized as the difference between high-quality and low-quality ones in our analysis. The evidence indicates a noteworthy variation in the estimated latent series along with the notable contrasts. The general thrust of the intertemporal comparison emphasizes a substantial deterioration in the estimated institutional quality in the post-2007 period. For instance, the series on voice and accountability appears to be close to zero residual threshold in the initial year of investigation (i.e. 1996) whilst it tends to deteriorates considerably after 2007. A similar pattern emanates from the estimated series on residualised political stability and absence of violence amid a sharp drop in the post-2007 period. Not all series tend to exhibit a path of pervasive decline in institutional quality. For instance, the residualized series on regulatory quality and government effectiveness uncover a pattern of accelerated growth after 2013 indicating some improvement in the efficacy of the public administration and civil service as well as some improvement in economic freedom. Yet, end-of-sample MCMC estimates are below the zero threshold indicating more exclusionary institutional structure compared to the expected level in geographically similar regions in the control sample. Furthermore, the estimated residual series for the rule of law and control of corruption indicate persistent and irreversible decline since the beginning of our sample. Such pattern indicates continuous deterioration in the effectiveness of legal institutions and ability to control corruption that unfolded shortly after the dissolution of Soviet Union.

FIGURE 2 – INSERT ABOUT HERE

Figure 3 exhibits the institutional quality trajectory for each dimension per individual regions in Ukraine.[17] The patterns of institutional quality development emphasize a general notion of diverse contrasts and notable similarities in the post-1996 institutional development. It may be said that some regions tend to have simulated latent institutional quality trajectory above the zero-residual threshold, indicating somewhat better institutional quality compared to their pre-determined exogenous geographic characteristics. For instance, the provinces such as Donetsk and Dnipro are characterized by relative stable institutional development until the 2014 Maidan revolution whilst they deteriorate afterwards with respect to residualized voice and accountability and political stability. Other regions such as Kharkiv exhibit a pattern of sharp deterioration in government effectiveness after 2007 and notable growth after 2014. The series on regulatory quality for Lviv further uncovers the presence of relatively more exclusionary economic institutions throughout the full period of investigation given that both the mean MCMC estimate and the respective confidence bounds are below zero. Despite some improvement after 2014, the mean point estimate fails to exceed the zero threshold indicating the continued persistence of the exclusionary economic institutions. A similar notion can be invoked for a region such as Odessa with respect to the rule of law. In particular, after a period of stable zero-residual series until early 2010, a sharp deterioration is perceptile afterwards amid a shift towards the strength of the rule of law that is considerably worse than expected from Odessa's pre-determined geographic characteristics. Furthermore, we also uncover the evidence of widespread weakness in the ability to control corruption in the capital city (Kyiv) that has remained steady in the post-independence period.

FIGURE 3 – INSERT ABOUT HERE

## 4 Identification Strategy

### 4.1 Setup

---

[17] The regions were randomly selected for comparison for each institutional quality dimension. A full and more exhaustive depiction is available upon request.

Our goal is to examine the contribution of deep institutional integration to the governance and institutional quality consistently. To this end, our aim is to estimate the appropriate counterfactual governance quality scenario to elicit the institutional quality cost of staying out of the deep institutional integration.

Suppose we observe a finite set of regions $(J + 1) \in \mathbb{N}$ over $T \in \mathbb{N}$ periods where $t = 1,2 \dots T$. An entry into the institutional integration that entails the characteristics of the treatment-based policy shock occurs at time $T_0$ and begins in $T_0 + 1$ and lasts until the end of the time period without interruption so that $t < T_0 < T$ and $T_0 \in \{1, T\} \cap \mathbb{N}$ (Abadie 2021).

Let $q_{j,t}^N$ be the potential governance outcome in j-th region in the hypothetical absence of the integration for $j \epsilon \{1, \dots J + 1\}$ and $t \in \{1, \dots T\}$, and let $q_{j,t}^I$ represent the observed realization of the governance outcome in the full period $t = 1,2, \dots T$. Without the loss of generality, the average governance effect of the institutional integration is defined as:

$$\alpha_{j,t} = q_{j,t}^I - q_{j,t}^N \qquad (6)$$

where $\alpha_{j,t}$ captures the difference between the observed realization of the governance outcome and the potential governance in the hypothetical absence of the institutional integration. Since the entry into the institutional integration can be as a binary variable switching between 0 and 1 and sub-periods $t < T_0$ and $t \geq T_0$, the entry into the integration can be described as a dummy variable that takes the value $D_{j,t} = 1 \; \forall \; \{j = 1\}$ and $D_{j,t} = 0$ otherwise. The effect of the entry into the institutional integration can be written as follows:

$$q_{j,t} = q_{j,t}^N + \alpha_{j,t} \cdot D_{j,t} \qquad (7)$$

The major identification constraint in estimating $\alpha_{j,t}$ arises from the fact that $q_{j,t}^N$ is unobserved to the econometrician and therefore has to be estimated to gauge the effect of staying out of the institutional integration. In this respect, our aim is to estimate the full vector of post-treatment effects of institutional integration $(\alpha_{1,T_0}, \dots \alpha_{1,T})$. Let $q_{j,t}^N$ be approximated through a simple latent factor model:

$$q_{j,t}^N = \delta_t + \boldsymbol{\theta}_t \cdot \mathbf{Z}_{j \in J} + \boldsymbol{\lambda_t} \cdot \boldsymbol{\mu}_{j \in J} + \varepsilon_{j,t} \tag{8}$$

where $\delta_t$ denotes the full set of time-fixed effects that absorb time-varying technology shocks common to all regions with constant loading, $\mathbf{Z}$ is a simple $(m \times 1)$ vector of observed auxiliary covariates unaffected by the entry into institutional integration, $\boldsymbol{\theta}_t$ is $(1 \times m)$ vector of prior unknown parameters, $\boldsymbol{\lambda_t}$ is $(1 \times H)$ vector of observed common factors, and $\boldsymbol{\mu}_j$ is $(H \times 1)$ a vector of unknown factor loadings, and $\varepsilon$ is the set of region-level transitory shocks under $\varepsilon \sim i.i.d$ structure (Xu 2017)

As advocated by Firpo and Possebom (2018), let $\mathbf{Y}_{j,t} = [Y_{j,1}, \ldots Y_{T_0}]$ be a vector of the observed realization of the outcomes for region $j \in \{1, \ldots J+1\}$ in the pre-intervention period where $t < T_0$, and let $\mathbf{X}_{j,t} = [X_{j,1}, \ldots X_{T_0}]$ be $[K \times 1]$ vector of covariates. Moreover, let $\mathbf{Y}_{j,t} = [\mathbf{Y}_2, \ldots \mathbf{Y}_{J+1}]$ be a matrix with $[T_0 \times J]$ dimension, and $\mathbf{X}_{j,t} = [\mathbf{X}_2, \ldots \mathbf{X}_{J+1}]$ a corresponding $[K \times J]$. Moreover, let $\mathbf{W} = (w_1, \ldots w_{J+1})$ be a simple vector of weights with $J \times 1$ dimension that captures the entire composition of the donor pool. Each particular value of the weight vector captures the weighted average of the donor pool's characteristics that best reproduce the institutional quality trajectory prior to the institutional integration implied from $\mathbf{Y}_{j,t}$ and $\mathbf{X}_{j,t}$ matrices. Thus, the fitted value of the outcome variable constructed from the characteristics of the donor pool for the given $\mathbf{W}$ is as follows:

$$\sum_{j=2}^{J+1} w_j \cdot q_{j,t} = \delta_t + \theta_t \cdot \sum_{j=2}^{J+1} w_j \cdot Z_j + \lambda_t \cdot \sum_{j=2}^{J+1} w_j \cdot \mu_j + \sum_{j=2}^{J+1} w_j \cdot \varepsilon_{j,t} \tag{9}$$

which implies that for each $t \in \{1, \ldots T\}$, we estimate $\hat{q}_{i,t}^N = \sum_{j=2}^{J+1} \widehat{w}_j \cdot q_{j,t}$. The optimal weights are derived by dividing the pre-treatment period training and validation sub-periods. which consists of the training and validation period. In the training period, the relative importance of covariates and pre-treatment outcomes is identified through a diagonal matrix $\widehat{\mathbf{V}}$ denoting the normalized variable weights. In the validation period, the weighing vector $\widehat{\mathbf{W}} = [\widehat{w}_1, \ldots \widehat{w}_{J+1}] \in \mathbb{R}^J : w_j \geq 0$ for each $j \in \{2, \ldots J+1\}$ captures the relative importance of each region in the loci of treated region's convex hull where (Botosaru and Ferman 2019)The set of weights selected on the basis of the similarity between $J = 1$ and $j \in \{2, \ldots J+1\}$ is given as a closed-form solution to the nested minimization problem (Becker and Klößner 2018):

$$\widehat{\mathbf{W}}(\mathbf{V}) = \underset{\mathbf{W} \in \mathbb{W}}{\mathrm{argmin}}(\mathbf{X}_1 - \mathbf{X}_0\mathbf{W})'\mathbf{V}(\mathbf{X}_1 - \mathbf{X}_0\mathbf{W}) \qquad (10)$$

Where $\mathbb{W} = \left\{ W = [w_2 \ldots w_{J+1}]' \in \mathbb{R}^J : w_J \geq 0 \text{ for each } j \in \{2, \ldots J+1\} \text{ and } \sum_{j=2}^{J+1} w_j = 1 \right\}$ and $\mathbf{V}$ is a diagonal positive semi-definite matrix having $K \times K$ dimension with a trace equal to one:

$$\mathbf{V} = \underset{\mathbf{V} \in \mathbb{V}}{\mathrm{argmin}} \left( \mathbf{X}_1 - \mathbf{X}_0\widehat{\mathbf{W}}(\mathbf{V}) \right)' \left( \mathbf{X}_1 - \mathbf{X}_0\widehat{\mathbf{W}}(\mathbf{V}) \right) \qquad (11)$$

Where notice that **W** is a weighing vector measuring the relative importance of weights from $j \in \{2, \ldots J+1\}$ potential sequence in the donor pool in the composition of synthetic control group, and **V** measures the relative predictive importance of each of K covariates and pre-$T_0$ outcomes (Billmeier and Nannicini 2013). Both weighing vector and covariate-level vector track and reproduce the institutional quality trajectory of the treated region as closely as possible. Hence, by choosing an appropriate distance matrix such as Euclidean or Trigonometric, the relative discrepancy between the treated region and its synthetic control group can be evaluated accordingly (Doudchenko and Imbens 2016). Without the loss of generality, the treatment effect of institutional integration on the institutional quality for $J = 1$ region and each $t \in \{1, \ldots T\}$ can be written as:

$$\lambda_{J=1,t} = q_{1t} - \sum_{j=2}^{J+1} w_j^* \cdot q_{j,t} = q_{1t} - \hat{q}_{1t}^N \qquad (12)$$

### 4.2 Inference

To evaluate the significance of the effect of institutional integration, our approach relies on the standard permutation test. In this respect, the question we ask is whether the estimated institutional quality gap is obtained by chance. Following Abadie and Gardeazabal (2003), Bertrand et. al. (2004), Abadie et. la. (2010), McClelland and Gault (2017), Firpo and Possebom (2018), we perform a series of in-space placebo simulation by iteratively applying the synthetic control estimator to the full set of regions that have joined the European Union to tackle the significance of the estimated effect of staying out of the European Union. The intuition behind the placebo simulations is both simple and straightforward. If the permutation of staying out of the European Union to the unaffected regions generates gaps of the level similar to the ones estimated for Ukrainian regions, our interpretation would be that the synthetic control analysis does not provide evidence of the significant effect of staying out of the European Union. By contrast, if the placebo simulation creates the institutional quality gaps across Ukrainian regions that are unusually large relative to the gaps for the regions that did not stay out of the European Union during the central and eastern European enlargement in 2004[18], 2007[19] and 2013[20], then the notion of the significant effect of staying out of the European Union becomes more plausible. To assess the significance of the estimated gaps, the synthetic control estimator is iteratively applied to every other region except Ukrainian ones which effectively shifts the latter from the treatment set to the donor pool. Henceforth, we compute the estimated effect for each placebo simulation which provides us the distribution of the estimated gaps for the regions that have not stayed out of the European Union.

Our approach is based on the benchmark for small-sample inference similar to Fisher's exact hypothesis test. Under such test, the stay-out of the European Union is permuted to the unaffected regions $j \in \{2, \dots J+1\}$ for each $t \in \{1, \dots T\}$. Therefore, for each $j \in \{2, \dots J+1\}$, $\hat{\lambda}_{j,t}$ is estimated for the sub-periods $t < T_0$ and $t \geq T_0$. In the next step, the full vector of post-treatment effects of staying out of the European Union, $\hat{\boldsymbol{\lambda}}_1 = [\hat{\lambda}_{1,T_0+1} \dots \hat{\lambda}_{1,t}]'$ is compared with the empirical distribution of $\hat{\boldsymbol{\lambda}}_j = [\hat{\lambda}_{j,T_0+1} \dots \hat{\lambda}_{j,t}]'$ based on the treatment permutation procedure. The notion behind the comparison yields a simple decision. If the vector of estimated effects for the treated regions is relatively large compared to the vector of effects for quasi-treated regions, the null hypothesis of no effect whatsoever of staying out of the European Union can be rejected.

A major caveat behind the comparison of $\hat{\boldsymbol{\lambda}}_1$ and $\hat{\boldsymbol{\lambda}}_j$ is that $|\hat{\lambda}_{1,t}|$ can be abnormally large in comparison with the empirical distribution of $|\hat{\lambda}_{j,t}|$ for some periods within $t \in \{1, \dots T\}$ but not for others. To partially account for the imperfect pre-$T_0$ fit of the institutional quality trajectory, we construct an empirical distribution of the root mean square error as a summary statistics:

$$RMSE_j = \frac{\sum_{t=T_0}^{T}(q_{j,t} - \hat{q}_{j,t}^N) \div (T - T_0)}{\sum_{t=1}^{T_0}(q_{j,t} - \hat{q}_{j,t}^N) \div (T_0)} \qquad (13)$$

---

[18] Czech Republic, Estonia, Hungary, Latvia, Lithuania, Poland, Slovakia, Slovenia
[19] Bulgaria and Romania
[20] Croatia

where $\sum_{t=1}^{T_0}(q_{j,t} - \hat{q}_{j,t}^N) \div (T_0)$ denotes pre-$T_0$ mean square predictive discrepancy between $J = 1$ and $j \in \{2, ... J + 1\}$, and $\sum_{t=T_0}^{T}(q_{j,t} - \hat{q}_{j,t}^N) \div (T - T_0)$ represents post-$T_0$ predictive discrepancy between the treated region and its synthetic peer. To determine whether the estimated gap between $q_{j,t}$ for $J = 1$ and $\hat{q}_{j,t}^N$ is statistically significant, we compute the two-sided p-value based on the treatment permutation procedure:

$$p = \frac{\sum_{j=1}^{J+1} \mathbb{I} \times (RMSE_j \geq RMSE_1)}{J+1} \qquad (14)$$

Where $\mathbb{I}(\cdot)$ is the indicator function revealing whether $RMSE_j$ from permuted treatment is, in absolute terms, larger than the benchmark RMSE of the treated region ($J = 1$). Without the loss of generality, the distribution of in-space placebos can be described as $\hat{\lambda}_{1t}^{Placebo} = \{\hat{\lambda}_{j,t} : j \neq 1\}$ and $p = Pr(|\hat{\lambda}_{1t}^{Placebo}| \geq |\hat{\lambda}_{1t}|) = \sum_{j \neq 1} 1 \times (|\hat{\lambda}_{1t}| \geq |\hat{\lambda}_{jt}|) \div (J)$. When the treatment is randomly assigned, the p-values may be interpreted through a classical randomization inference. By contrast, if the treatment is not randomly assigned, the p-value may be interpreted as the proportion of quasi-treated regions that has an estimated institutional quality gap at least as large as the treated region. By inverting the p-values in the intertemporal distribution, the confidence bounds based on the pre-specified significance threshold can be constructed henceforth. Under the null hypothesis, $H_0: q_{j,t} = \hat{q}_{j,t}^N$ for each $j \in \{1, ... J + 1\}$ and $t \in \{1, ... T\}$. Since our donor pool is reasonably large, our benchmark rejection rule stipulates exact null hypothesis of no effect whatsoever where the p-values are constructed by inversely weighing the estimated placebo gaps:

$$p = \sum_{j=1}^{J+1} \pi_j \times \mathbb{I}[RMSE_j \geq RMSE_1] \qquad (15)$$

where $\pi$ denotes the analytical weight in $j \in \{2, ... J + 1\}$ placebo set where placebos with smaller magnitude pre-$T_0$ $RMSE$ than that of $J = 1$ are overweighed to avoid obtaining the p-value driven by the extreme relative rarity of obtaining a large effect from poorly fit placebos. As noted by Abadie et. al. (2010, p. 502), placebo runs with a poor quality of the fit do not provide a meaningful information to measure the relative rarity of estimating a large post-$T_0$ gap for the unit with a good quality of fit prior to the treatment. For this reason, $\pi$ weighs the placebos based on the size of $RMSE$ and overweighs those with the quality of fit closer to $J = 1$ to avoid artificially low p-values as a result of poorly fitted placebos.

# 5    Results

## 5.1    Baseline results

Table 1 reports the pre-2007 imbalance in institutional quality outcomes and the auxiliary covariates for each individual outcome and selected provinces. The descriptive evidence suggests that the synthetic control estimator through the latent factor model from Eq. (2) provides an excellent fit of the institutional quality trajectory of between the Ukrainian regions and their synthetic peers before 2007 based on the weights obtained in training and validation period. In each respective specification, the size of RMSE appears to be very low and within the conventional 1% error discrepancy when compared against the zero-fit benchmark model (Adhikari and Alm 2016). This suggests that pre-treatment imbalance in covariates and outcomes is somewhat unlikely to taint the effect of staying out of European Union on the institutional quality of Ukrainian provinces.

TABLE 1 – INSERT HERE

Figure 4 presents the average treatment effect of staying out of European Union for the full province-level treatment sample. The evidence invariably indicates large and persistent institutional quality losses from staying out of the European Union. In particular, the synthetic control estimates suggest that Ukrainian provinces have developed substantially worse institutional quality by staying out of the European Union compared to the regions and counties from the countries that have entered the European Union in 2004 and later. Although the estimated set of average treatment effects indicates invariably worse trajectory of institutional quality after 2007, a notable outcome-specific heterogeneity. More specifically, the largest drop in institutional quality outcome is perceptible for the level of political stability and absence of violence. The estimated average treatment effect on the treated (ATET) is -0.39 basis points which implies that Ukrainian provinces political stability in response to post-2007 is around 40 percentage points lower than it would be if Ukraine were hypothetically admitted to the European

Union at the same time as the rest of central and eastern Europe. In a similar vein, the estimated ATET of staying out of the European Union on the ability to control corruption is -0.21 basis points which is around half as large as the effect on political stability. It invariably indicates a widespread deterioration of the control of corruption after 2007 compared to Ukrainian central and eastern European peers that were admitted to the EU. Hence, our estimates suggest a marked improvement in the ability to control corruption if Ukraine were hypothetically admitted to the European Union in the same vein as its central and eastern European control group from 2004 and subsequent rounds of enlargement. In a similar vein, the estimated ATET with respect to the rule of law is around -0.15 basis points and confirms large and negative effects of staying out of the European Union on the strength of the rule of law. The behaviour of ATET with respect to the control of corruption and rule of law is particularly informative. It indicates a marked and pervasive drop in the observed series after 2007 amid a slight but slow increase after 2014. Our estimates thus indicate somewhat improvement in the strength of rule of law and ability to control corrupt behaviour after the Maidan Revolution although the magnitude of the closing of the gap is small compared to the speed of the decline after 2007. A similar pattern is perceptible with respect to the quality of regulation. In particular, the estimated ATET is around -0.14, indicating a large drop in regulatory quality after 2007. By contrast, the magnitude of the drop in the residualized voice and accountability is small and around 0.07 basis points in the post-treatment period. Similar to the patterns found elsewhere, voice and accountability series exhibit a tendency of recovery in the years after Maidan Revolution although the gap appears to be far from recovery compared to pre-2007 peak. In stark contrast, the effect of staying out of the European Union on the effectiveness of government administration appears to be temporary. Unlike the evidence uncovered for the five other dimensions of institutional quality, government effectiveness exhibits a sharp drop after 2007, and a rapid increase and recovery after 2014. The end-of-sample ATE estimate indicates a full recovery of government effectiveness compared to the pre-2007 peak level. This implies that despite staying out of the European Union, government effectiveness and quality improved substantially and to the same degree as in the central and eastern European synthetic control groups of Ukrainian regions. On balance, the estimates based on our analysis uncover evidence of the pervasive breakdown of institutional quality beginning to unfold in 2007 with respect to the political stability, control of corruption, level of liberal democracy, rule of law and regulatory quality. By contrast, staying out of the European Union induced only a temporary deviation of the residual government effectiveness series from its pre-2007 equilibrium. From the normative perspective, our estimates imply that a hypothetical admission of Ukraine to the EU would substantially improve the institutional quality whereas the admission to the EU does not appear to be a necessary condition for improving the institutional quality of government administration where plausible policy alternatives can equally improve the efficacy of the administration. Given previously recognized importance of the rule of law, political stability and control of corruption for sustained economic growth and social development, our conjecture is that the admission into a more inclusive institutional integration compared to Russian institutional orbit would generate marked, substantial and permanent gains in institutional quality.

FIGURE 4 – INSERT HERE

Figure 5 reports the estimated institutional quality effect of staying out of the European Union for selected Ukrainian provinces in the 1996-2020 period. Our empirical strategy is able to almost perfectly reproduced the pre-2007 trajectory of residualised institutional quality of Ukrainian provinces from the convex combination of the intrinsic attributes and characteristics of central and eastern European regions which effectively renders RMSE parameter close to zero. Without the loss of generality, province-level evidence is fully consistent with the average region-level treatment effect of staying out of the European Union. In particular, the provincial institutional quality trajectory appears to exhibit a structural breakdown after 2007 indicated by permanently lower level of quality in comparison with the counterfactual scenario based on the admission to the European Union in 2007. The only exception to the structural breakdown emanates from the post-2007 effect on government effectiveness where the treated region (i.e. Zaporizhznia) tends to gradually close the gap induced by the post-2007 instability behind the central and eastern European peers. The evidence suggests that some regions tend to have a large gap behind their respective synthetic control group. For instance, the ability to control corruption in Kyiv City permanent deteriorates from its synthetic peer, yielding a large quality gap despite some improvement after 2014.[21]

FIGURE 5 – INSERT HERE

Figure 6 reports the frequency distribution of synthetic control groups. Given the size of the treatment sample and the donor pool, our preferred approach is to report the frequency of donor regions in the synthetic control groups of Ukrainian provinces to elicit the most powerful and important donor regions with non-zero weight in the synthetic control group of Ukrainian provinces. Notwithstanding the contrasts, notable variation in the composition of synthetic control groups exists across the treated provinces. For instance, Kyiv City's trajectory of the residual control of corruption prior to 2007 can be best reproduced as a convex combination of the regions whose implied attributes fall within the convex hull of the respective city such as Bihor [Romania] (30%), Riga [Latvia] 16%, Satu Mare [Romania] 15%, Krapina-Zagorje [Croatia] (12%), Zilinsky Kaj [Slovakia] 9%, Maramures [Romania] (7%), Medjimurje [Croatia] 7%, Sofia Stolitsa [Bulgaria] 3%, and Bucharest [Romania] (2%). In terms of further example, Kharkiv's political stability and absence of violence trajectory can be best reproduced as a weighted combination of the implied institutional quality attributes and its exogenous characteristics of Heves [Hungary] 35%, Pernik [Bulgaria] 22%, Szabolcz [Hungary] 11%, Jihomoravsky kraj [Czech Republic] 8%, Carinthia [Slovenia] 7%, Zilinsky kraj [Hungary] 7%, Pardubicky kraj [Czech Republic] 6%, Trnavsky kraj [Slovakia] 4%, respectively. Furthermore, Odessa's residual rule of law trajectory can be best reproduced as a weighted combination of the implied institutional quality attributes and auxiliary characteristics of Gorj [Romania] 61%, Zala [Hungary] 11%, Ogres [Latvia] 9%, Liberecky kraj [Czech Republic] 9%, Varazdin [Croatia] 5%, Prague [Czech Republic] 3%, Veliko Tarnovo [Bulgaria] 3%. From a more general perspective, the synthetic control groups of the treated Ukrainian provinces in the voice and accountability specifications are dominated by Prague and Medjimurje followed by a the attributes of Czech regions and several Bulgarian and Hungarian regions. This implies that a convex combination of attributes consists of Czech regions, Croatian counties and several other central and eastern European regions is very well able to reproduce the observed dynamics of voice and accountability prior to the year of hypothetical admission to the European Union. In the specifications of political stability and absence of violence models, the synthetic control groups are dominated by Medjimurje [Croatia], Lubuskie [Poland] and Zilinsky kraj [Slovakia]. Other important donors with non-zero weight include the capital cities in central Europe such as Prague and Zagreb as well as several regions in eastern Europe, especially Romania and Bulgaria, such as Constanta, Montana and Bistrita-Nasaud. By contrast, the synthetic control groups in the specifications using the government effectiveness as the outcome of interest are disproportionately dominated by Romanian regions followed by Slovak and Polish regions whilst other regions enter the control groups considerably less frequently. In addition, the synthetic control groups in the regulatory quality specifications appear to be most homogenous given a considerably smaller number of donors with non-zero weight. The synthetic control groups are dominated by Prague, Sofia, three Croatian counties (Krapina-Zagorje, Medjimurje, Koprivnica-Krizevci) and one Polish regions (Kujawsko-Pomorskie). In a similar vein, the rule of law trajectories of Ukrainian provinces prior to 2007 can be most eloquently reproduced as a weighted combination of the quality-related and auxiliarly attributes dominated by Croatian counties, Czech regions and Bulgarian provinces whilst others enter the groups much less frequently. An interesting insight emanates from the donor frequency comparison in the specifications involving the control of corruption as the underlying outcome. The synthetic control groups disproportionately consist of the Zagreb City whose additive weight almost equals the sum of weights for the other donor regions. Moreover, the synthetic control groups reflecting the effect of staying out of the European Union are also characterized by the presence of several Romanian regions, Latvian and Lithuanian districts.

---

[21] For the sake of the size of our treatment sample and space limitations, the figures report the average treatment effect only for selected regions. An exhaustive report for the full set of regions or for alternative combination of regions is available upon request.

FIGURE 6 – INSERT HERE

*5.2    In-space placebo analysis*

Our approach to tackle the statistical significance of the effect of staying out of the European Union on the institutional quality of Ukrainian provinces consists of the in-space placebo analysis. In particular, the stay-out of the European Union is assigned to the regions in the donor pool that have not stayed out of the European Union which effectively shifts Ukrainian provinces to the donor pool. Through an iterative application of the synthetic control estimator to the unaffected regions, we estimate a large vector of placebo gaps and evaluate the estimated effects for Ukrainian provinces compared to the effect sequence from placebo runs. If the institutional quality effects estimated for Ukrainian provinces are particularly large and unique, the notion of significant effects of staying out of the European becomes more credible. By contrast, if the gaps estimated for Ukrainian provinces do not seem to be substantially different from the placebo gaps, then our conclusion is that the analysis fails to provide evidence of significant effect of staying out of the European Union. For each institutional quality outcome under consideration, we compute two-tailed p-values as $p = Pr(|\hat{\lambda}_{1t}^{Placebo}| \geq |\hat{\lambda}_{1t}|) = \sum_{j \neq 1} 1 \times (|\hat{\lambda}_{1t}| \geq |\hat{\lambda}_{jt}|) \div (J)$ and proceed with the benchmark rejection rule imply no effect whatsoever through inversely weighted placebo gaps. Given the size of the treatment sample and the donor pool and the length of the pre-treatment period, we calculate more than 34 billion placebo gap averages. A relatively large number of placebo averages allows us to use almost full random sampling approach towards treatment permutation and improves the consistency of the estimated p-values denoting the probability of no effect whatsoever. Apart from the standard p-values, we also undertake a simple difference-in-differences analysis of placebo gap coefficients to determine parametrically whether the gaps estimated for Ukrainian provinces are statistically significantly different from the placebo gaps. Each specification also contains the full set of region-fixed effects and time-fixed effects to absorb the confounding influence of the heterogeneity bias from the estimated post-treatment coefficient.

Table 2 reports in-space placebo analysis of staying out of the European Union. Our key parameter of interest is the $\lambda_{k=1,full}$ which denotes the average post-2007 institutional quality gap across Ukrainian provinces for each institutional outcome under consideration. The evidence from the placebo analysis uncovers a high degree of uniqueness of the institutional quality gap. Controlling for the unobserved region-specific and time-varying heterogeneity, the estimated coefficients for the Ukrainian province-level post-treatment gaps are large and statistically significant at the 1 percent level. Consistent with the estimated magnitudes, the largest post-treatment coefficients are found with respect to the political stability and absence of violence, control of corruption and the strength of the rule of law. The difference-in-differences analysis thus largely confirms the significance of the estimated gaps for Ukrainian provinces and highlights relatively high losses of staying out of the European Union.

TABLE 2 – INSERT HERE

Figure 7 reports the p-values on the institutional quality effect of staying out of the European Union through the simulation-based placebo analysis. The figure depicts the proportion of regions in the donor pool having at least as large effect of staying out of the European Union as Ukrainian regions. Thus, a high proportion indicates considerably weaker effect of stay-out whilst a low proportion corresponds to statistically more significant effect of staying out. More specifically, the figure does not report an average p-value from the placebo runs but the exact p-values on no effect whatsoever for each post-treatment period and each respective institutional quality outcome. Without the loss of generality, the p-values may be used to conduct inference on the institutional quality effect of remaining under Russian institutional and political patronage. The evidence suggests unequivocally low p-values associated with the staying out of the European Union after 2007. In particular, the simulated p-values are around 0.000 threshold for the entire post-treatment period for series reflecting voice and accountability, political stability, rule of law and control of corruption. From the substantive perspective, the inference on the effect of staying out of the European Union implies that by hypothetically joining the EU, Ukrainian regions would develop superior rule of law, exceptional political stability and absence of violence and substantially stronger control of corruption along with a more liberal democratic governance compared to what they've developed by themselves under the Russian institutional and political influence. By contrast, the p-values on the quality of regulation do not indicate statistically significant effect of staying out of the European Union on the respective institutional quality. In terms of further example, the statistical significance of the effect on government effectiveness disappears almost entirely in the end-of-sample post-treatment period when the p-value on the stay-out effect moves from 0.04 to 0.68 which implies that the improvement in the quality and efficacy of government administration emanating from the EU institutional integration is long-lasting but temporary institutional quality shock. On balance, the improvements in the strength of anti-corruption legislation and enforcement, liberal democratic governance, freedom of the press, level of the rule of law and political stability invariably suggests that the institutional quality improvement outside the extractive institutional orbit is both large, statistically significant and outweighs the temporary nature of the effect on government effectiveness as well as the insignificant effect on the quality of regulation.

### 5.3    Transmission mechanisms at work

The key question behind the estimated institutional quality effects of staying out of the European Union concerns the transmission mechanisms at work which might explain how the failure to embark on the path of European integration translates into lower institutional quality than Ukrainian regions would have achieved otherwise. Our approach to tackle the transmission mechanisms considers both long-term historical legacies as well as more contemporaneous channels such as culture, state capacity, geographic and natural resource endowments, and level of economic development based on the region-level data provided in Gennaioli et. al. (2014, 2015).

Long-term historical legacies of institutional regimes from the past matter for economic and social development. A comprehensive study by Becker et. al. (2016a) emphasize a positive historical legacy of the Habsburg empire in the subjected territories in terms of higher level of trust of citizens in local public services. Using a geo-referenced border specification within two-dimensional regression discontinuity design (RDD), their findings suggest that historical Habsburg affiliation both increases the level of trust in government and is associated with lower corruption in courts and police. Similar evidence on the persistent and long-lasting effects of historical legacies associated with Habsburg empire is reported in Dimitrova-Grajzl (2007), Schulze and Wolf (2009), Grosjean (2011), Mendelski and Libman (2014), and Grosfeld and Zhuravskaya (2015) among several others.

The era of Habsburg empire in modern-day Ukraine lasted from 1772 to 1918. The Habsburg empire acquired a sizeable part of the Ukrainian territory in 1772 whilst beforehand only the Transcarpathia was integrated into the Habsburg empire. Upon the first partition of Poland in 1772, the Habsburgs acquired the territories of Galicia and Lodomeria. Three years later, the small territory of Bukovina was acquired from the Ottoman Empire and integrated into the Habsburg empire. If the Habsburg legacy generates a positive effect on the trust in public institutions, the question that remains is whether formerly Habsburg territories exhibit disproportionately larger losses of staying out of the European Union. Given that our observations are disaggregated to the regional level, we cannot perform an RDD since it would require individual or spatially more disaggregated data for such purpose. However, given the intrinsic sample size limitations, we assess whether the Habsburg legacy explains the contrasts in the benefits and losses of staying out of the European Union by testing the equality of effects estimated through synthetic control estimator between the Ukrainian provinces integrated into the Habsburg empire (Magocsi 2002)[22] and those outside the empire.

Table 3 reports the tested equality of estimated institutional gaps for each outcome in the investigation before and after the date of hypothetical admission to the European Union. The table reports the average differences in the outcome-specific institutional quality gap between the provinces integrated in the Habsburg empire and those not integrated therein. The key parameter of interest is the difference in the effect coupled with the p-values on the test statistics. The evidence offers a limited empirical support for the persistent effect of Habsburg empire on institutional quality gaps. The provinces under former Habsburg rule tend to have somewhat lower effect on political instability and absence of violence, indicting relatively larger stability benefits of joining the EU for the regions that used to be under strong historical influence of Russian empire. Formerly Habsburg-ruled Ukrainian regions also have somewhat slower deterioration of government effectiveness and regulatory quality in response to staying out of the European Union, and marginally lower effect with respect to the rule of law. By contrast we find no evidence to corroborate the notion that belonging to the Habsburg empire entails substantially better control of corruption or a greater freedom of the press and more rigorous democratic accountability.

---

[22] Chernitvsi, Ivano-Frankivsk, Lviv, Ternopil, Zakarpattya

TABLE 3 – INSERT HERE

Figure 8 exhibits the estimated effects for the cultural, religious, geographic and economic mechanisms at work. The vertical axis depicts the first principal component of the estimated institutional quality gaps to provide a more comprehensive measure of loss with the largest common variation without losing either generality or specificity of the gap for each outcome. The evidence invariably suggests reasonably strong links between culture, physical geography and level of economic development in contrasting the estimated effects of staying out of the European Union. For instance, the notion that the exposure to Protestant religion bolsters economic growth through its strong adherence to work ethic and spirit of entrepreneurship initially emphasized by Weber (1904) has been a subject of rigorous scholarly debate among economists, sociologists and political scientists (Ekelund et. al. 2002, Becker et. al. 2016b). Some find evidence in favor of the Weber hypothesis (Grier 1997, Delacroix and Nielsen 2001, Becker and Woessmann 2009, Schaltegger and Torgler 2010, Bai and Kung 2015, Becker and Won 2021), others provide evidence indicating no long-term effects of Protestantism on economic outcomes (Doepke and Zilibotti 2005, 2008, Van Hoorn and Maseland 2013, Wang and Lin 2014, Cantoni 2015). We collect the data on the share of Protestants and Eastern Orthodox denominations for each Ukrainian province for the period of our investigation from Lyubitseva (2014), and find that provinces with higher share of Protestant religious denomination tend to have considerably lower losses of staying out of the European Union. Consistent with the prior literature on the economic and institutional quality effects of Protestantism, the negative relationship suggests that a somewhat stronger Protestant ethnic can partially compensate for the negative institutional quality effect of staying out of the European Union. By contrast, the provinces with a greater share of Eastern Orthodox denominational structure tend to have considerably higher institutional quality losses of staying out of the European Union, which readily suggests that a deep institutional integration based on relatively more inclusive institutional framework can disproportionately bolster quality improvements in the provinces with a large exposure to Eastern

Orthodox religious denomination. In addition, the provinces with a greater share of population with pro-Russian support tend to benefit somewhat less from being hypothetically admitted to the European Union compared to the provinces with a smaller share of pro-Russian supporters.

FIGURE 8 – INSERT AROUND HERE

In terms of further example, the provinces located further away from the coast tend to benefit substantially more from the EU-induced institutional quality improvement than those in a closer proximity to the coast. Consistent with the economic geography literature (Limão and Venables 2001), the costs of staying out of the European Union are somewhat higher for the inland and landlocked provinces that lack access to the international sea waters, higher transport costs and higher costs of trade (Frankel and Romer 1999). We also show that the exposure to malaria ecology (Sachs and Malaney 2002) adversely affects the institutional quality benefits of joining the European Union. Furthermore, provinces with a lower level of state capacity to provide public goods such as education, law and order tend to have somewhat higher losses of staying out of the European Union which is consistent with prior findings (Besley and Persson 2010, Acemoglu et. al. 2015, Dincecco 2017). In addition, provinces with higher level of economic development, proxied either by per capita GDP or per capita luminosity (Henderson et. al. 2012) tend to lose relatively less by staying out of the European Union compared to the poorer provinces. Taken together, this implies that a deep institutional integration would expectedly deliver stronger institutional quality improvements to the provinces with lower level of economic development. By contrast, the natural resource endowment, captured by the share of oil and gas production in the province-level GDP, only weakly correlates with the losses of staying out of the EU and do not seem to play a significant role.

## 6 Conclusion

In this paper, we examine the institutional quality effects of the institutional integration. To this end, we estimate the effects of staying out of the European Union on the institutional quality of the Ukrainian provinces for the period 1996-2020 by exploiting the political crisis in 20007 as a vacuum generating a potential admission to the European Union. By comparing the institutional quality trajectory of 24 Ukrainian provinces with the donor pool of 195 central and eastern European regions that were admitted to the European Union in 2004 and subsequent enlargement rounds, we assess the institutional quality cost of remaining under Russian political influence and highlight the potential benefits of the admission to the European Union. By making use of the novel dataset measuring subnational institutional quality generated through the application of Bayesian posterior analysis supported by machine learning-based algorithm, we are able to better unravel the heterogeneity in the benefits of the hypothetical European Union membership.

Through the application of synthetic control estimator, originally proposed by Abadie and Gardeazabal (2003) and extended by many other scholars, we find evidence of significant and substantial institutional quality improvements in response to the hypothetical admission of Ukraine to the membership in the European Union. The evidence invariably pinpoints substantial institutional quality gains in terms of improved political stability, reduced violence, strengthened liberal democratic governance, stronger rule of law and better control of corruption. In slight contrast, the improvements proceeding from improved regulatory quality and effectiveness of government administration seem to be temporary. Without the loss of generality, the evidence thus suggests that the hypothetical admission to the European Union can generate a positive deep institutional shock promulgating the improvements of deeper layers of institutional quality such as the rule of law which Ukraine can seldom develop from its domestic political economy equilibrium without the EU membership. By contrast, policy improvements leading to better regulatory design and enforcement of competition and structural policies do not necessitate deep institutional integration. Given the previously recognized primacy of rule of law and institutional instability for economic growth (Weingast 1995, Feng 1997, Rodrik et. al. 2004), the benefits of the deep institutional integration proceeding from the hypothetical membership in the European Union are almost impossible to be left to neglect.

Our analysis provides several noteworthy policy-relevant normative implications. First, by staying out of the European Union, Ukrainian provinces' institutional quality has deteriorated considerably compared to the trajectories in the donor pool of central and eastern European regions that have been integrated into the Western economic and political circuit of influence.

By staying under the contours of Kremlin's political influence and interference into domestic political economy, Ukrainian provinces have undergone a worsening control of corruption, a steady backdrop of the rule of law, widespread limitations on the freedom of express, and rampant deterioration of political stability. Furthermore, we tackle the statistical significance of the post-2007 institutional quality gaps through a large-scale placebo analysis that involves a simulation with more than 34 billion placebo averages for each institutional quality outcome under consideration. We find that in spite of the noteworthy institutional reforms that took place after 2018, the deterioration of institutional quality in post-2007 period appears to be permanent with the exception of the government effectiveness and regulatory quality. Second, policy efforts geared towards institutional reforms that address the negative bias proceeding from Kremlin's political influence and interference abroad may generate growth- and stability-enhancing effects provided that the de facto enforcement of such policies is reasonably strong. And third, despite the setbacks posited by Brexit and COVID19 pandemic, the European Union institutional framework provides a superior set of institutions and policies that the prospective candidates for membership cannot develop domestically given the capacity constraints, knowledge and coordination failures or any inherent domestic political economy bias.

As a final note, it should be stressed that additional checks of external validity are necessary before a definitive conclusion may be drawn on the generalization of the European Union's institutional integration and its effect on institutional quality and economic performance. Whilst our investigation casts significant benefits of the hypothetical membership for Ukraine, extending the scope of analysis to the regions with similar historical experience and contemporaneous domestic instability such as Western Balkans that can be considered at least remote candidates for EU membership would uncover interesting insights on the potential benefits of the admission that could be compared to the benefits estimated for Ukraine.

**Figure 1**: The structural break of 2007 political crisis and Ukraine's institutional quality trajectory

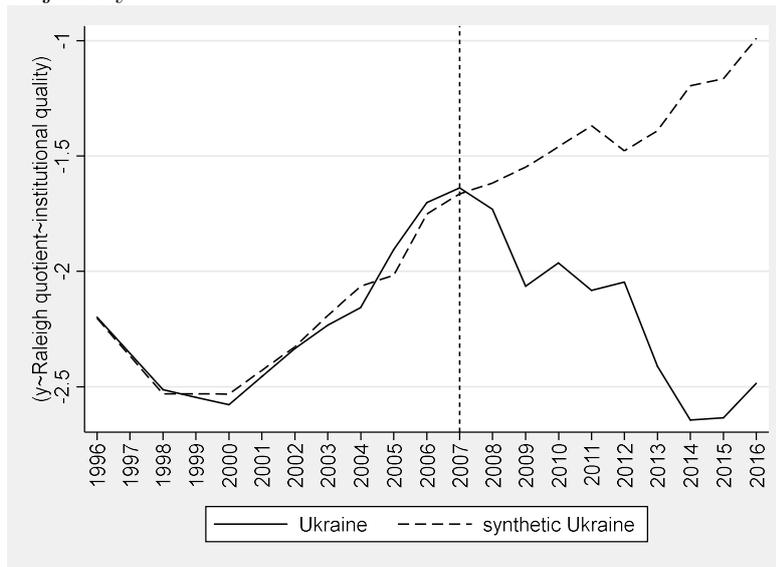

**Figure 2**: Institutional quality residuals across Ukrainian provinces, 1996-2020

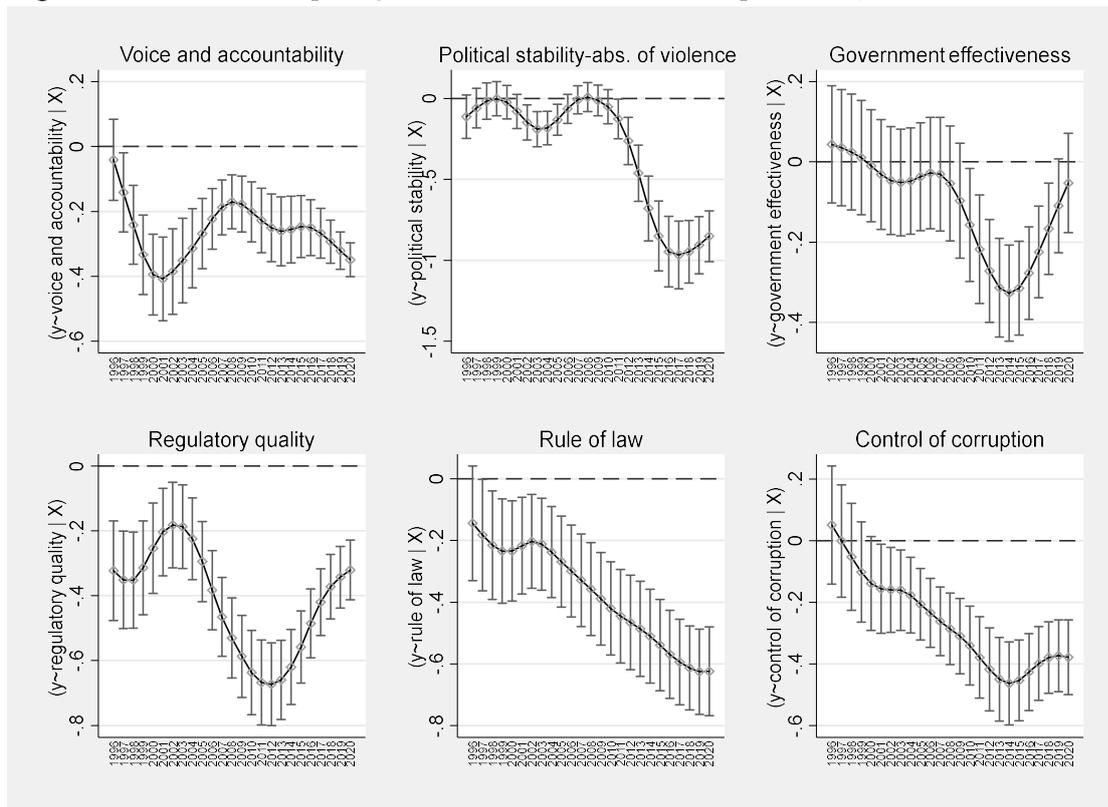

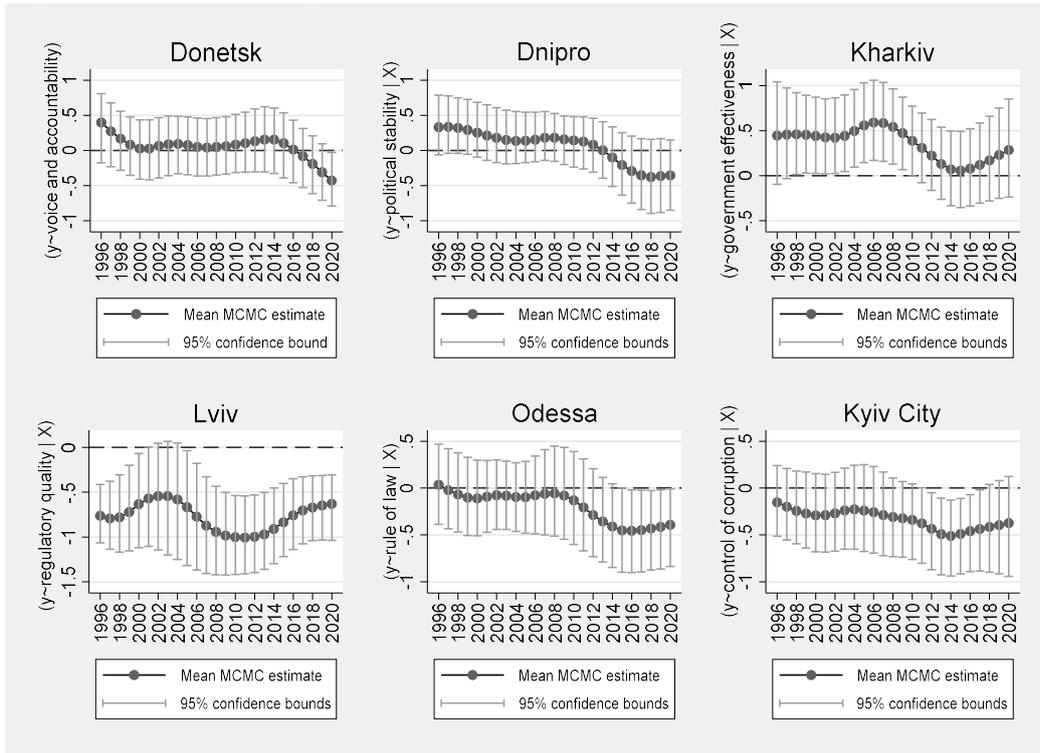

**Figure 3**: Institutional quality residuals across selected Ukrainian oblasts, 1996-2020

Table 1: Pre-treatment outcome and auxiliary covariate imbalance

| | Voice and accountability | | Political stability and absence of violence | | Government effectiveness | | Regulatory quality | | Rule of law | | Control of corruption | |
|---|---|---|---|---|---|---|---|---|---|---|---|---|
| | Zhytomyr | | Kherson | | Zaporizhzhya | | Luhansk | | Odessa | | Kyiv City | |
| RMSE | 0.015 | | 0.002 | | 0.001 | | 0.002 | | 0.003 | | 0.0024 | |
| | Treated | Matched | Treated | Matched | Treated | Matched | Treated | Matched | Treated | Matched | Treated | Matched |
| Outcome variable in 1996 | -0.29 | -0.27 | 0.08 | 0.08 | 0.57 | 0.57 | 0.19 | 0.19 | 0.04 | 0.03 | -0.15 | -0.15 |
| Outcome variable in 1998 | -0.38 | -0.41 | 0.08 | 0.08 | 0.44 | 0.44 | 0.15 | 0.15 | -0.07 | -0.07 | -0.24 | -0.24 |
| Outcome variable in 2000 | -0.53 | -0.52 | 0.03 | 0.03 | 0.31 | 0.31 | 0.20 | 0.20 | -0.11 | -0.11 | -0.29 | -0.29 |
| Outcome variable in 2002 | -0.52 | -0.50 | -0.07 | -0.07 | 0.20 | 0.20 | 0.20 | 0.20 | -0.08 | -0.08 | -0.26 | -0.27 |
| Outcome variable in 2004 | -0.41 | -0.42 | -0.08 | -0.08 | 0.18 | 0.18 | 0.14 | 0.14 | -0.09 | -0.09 | -0.23 | -0.23 |
| Outcome variable in 2006 | -0.35 | -0.33 | -0.16 | -0.06 | 0.20 | 0.20 | 0.03 | 0.03 | -0.078 | -0.08 | -0.26 | -0.26 |
| Latent institutional quality | -2.90 | -2.89 | -0.10 | -0.07 | 1.86 | 2.62 | 3.22 | 2.84 | -0.57 | -0.59 | -1.97 | -2.03 |
| Latitude | 50.25 | 46.71 | 46.63 | 46.55 | 47.83 | 51.82 | 48.56 | 49.84 | 46.48 | 46.88 | 50.45 | 48.68 |
| Longitude | 28.66 | 15.94 | 32.58 | 20.07 | 35.16 | 22.21 | 39.33 | 21.41 | 30.74 | 21.44 | 30.52 | 21.31 |
| Capital | 0 | 0 | 0.00 | 0.00 | 0.00 | 0.00 | 0 | 0 | 0.00 | 0.00 | 1 | 0.199 |
| Landlocked | 1.00 | 0.64 | 0.00 | 0.61 | 0.00 | 0.60 | 1 | 0.93 | 0.00 | 0.91 | 1 | 0.843 |
| Land area (log) | 10.30 | 7.18 | 10.26 | 8.86 | 10.21 | 9.07 | 10.19 | 9.43 | 10.41 | 8.29 | 9.51 | 8.20 |
| Altitude | 226.00 | 134.65 | 48 | 157 | 60.00 | 138.26 | 81.00 | 128.76 | 43 | 207.12 | 155 | 201.09 |
| Temperature | 6.80 | 11.72 | 10.3 | 10.1301 | 9.4 | 8.1216 | 8.80 | 9.00 | 10.7 | 9.8007 | 8.4 | 9.59 |
| Rainfall | 570.00 | 705.38 | 453 | 567.52 | 498 | 596.622 | 503 | 538 | 453 | 633.44 | 618 | 641.06 |
| Sunshine duration | 1843 | 2169 | 2217 | 2217 | 1978 | 1762 | 2009 | 1831 | 2183 | 2016 | 1843 | 1992 |

**Figure 4**: Average region-level treatment effect of staying out of European Union in Ukraine, 1996-2020

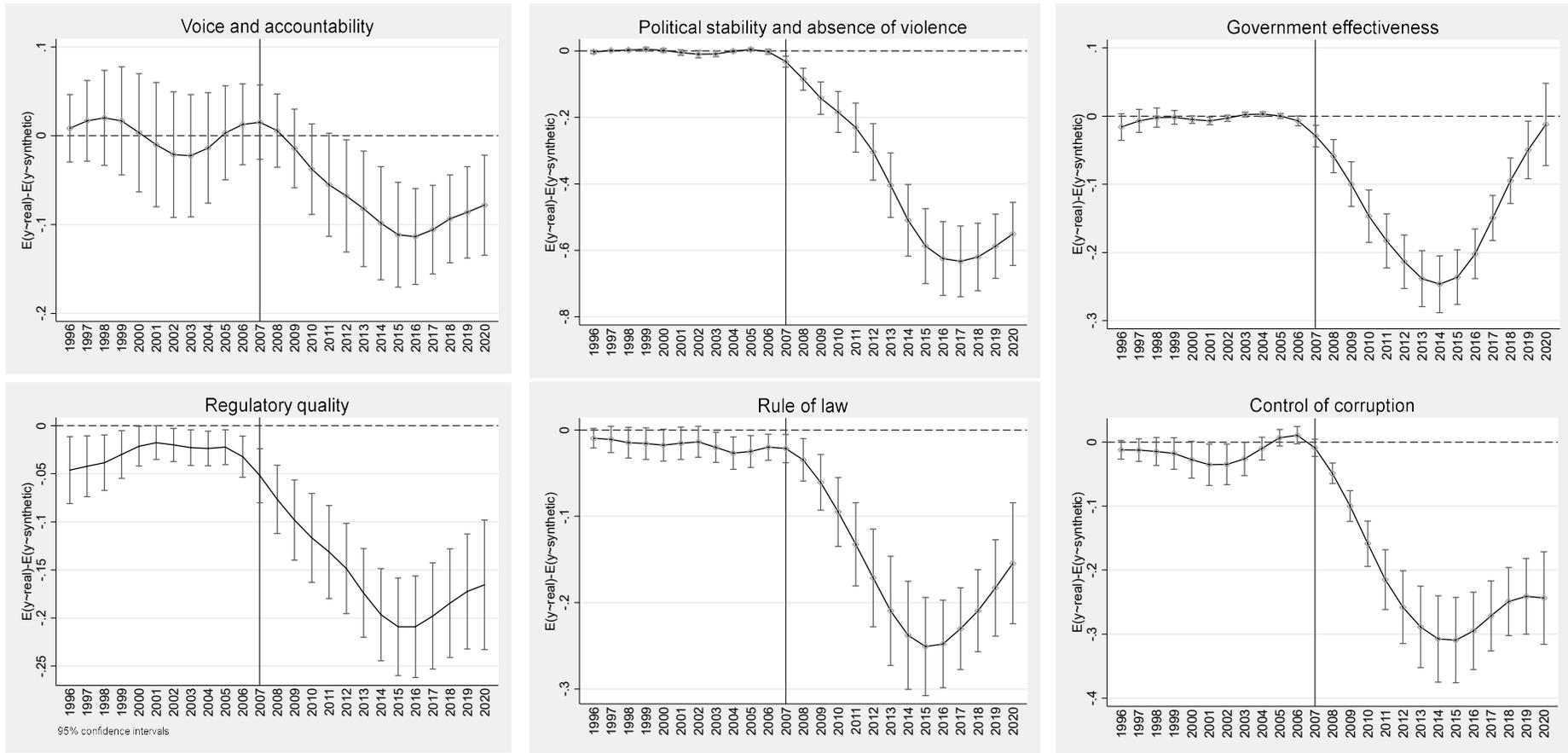

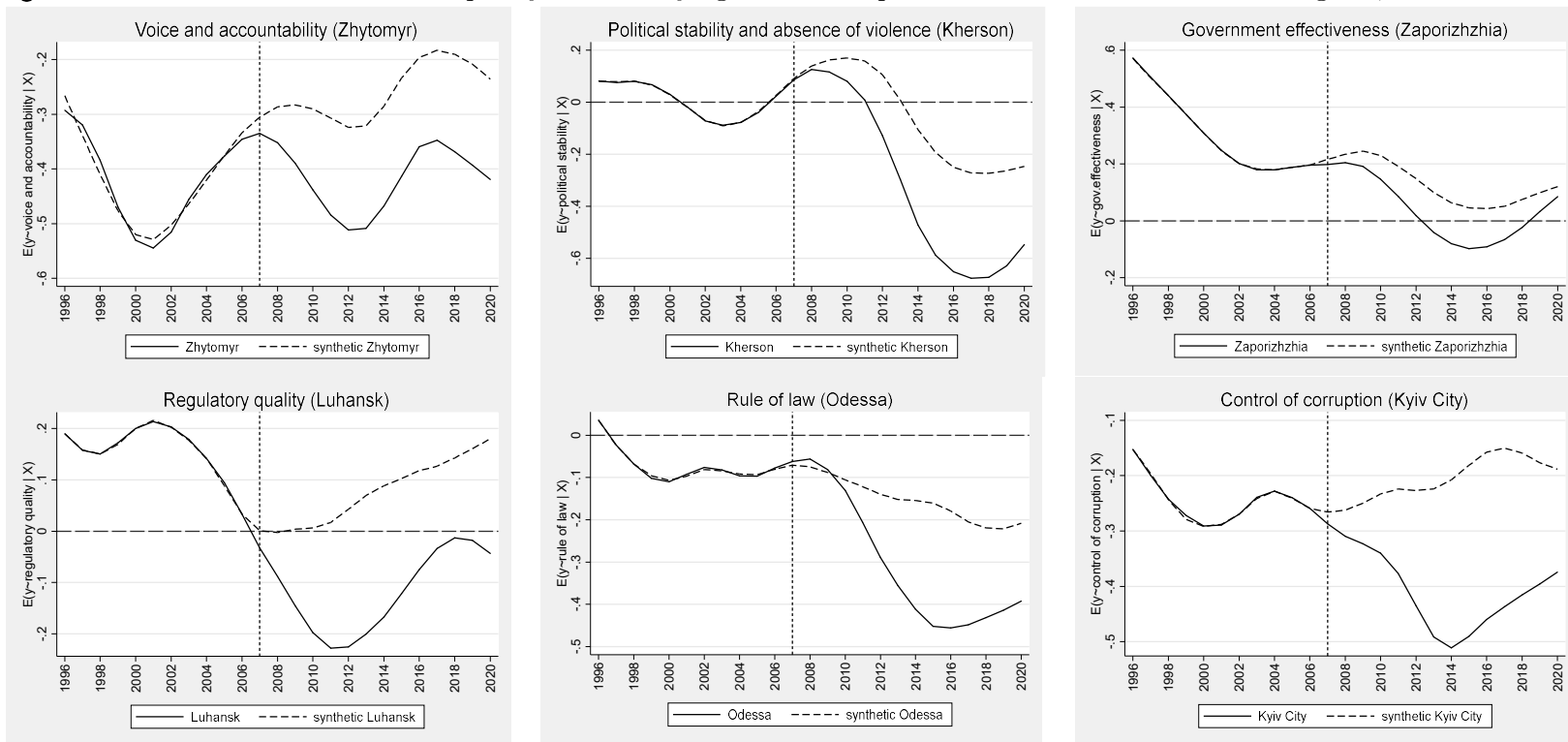

**Figure 5**: The estimated institutional quality cost of staying out of European Union in selected Ukrainian regions, 1996-2020

**Figure 6**: Frequency distributions of synthetic control groups for Ukrainian regions

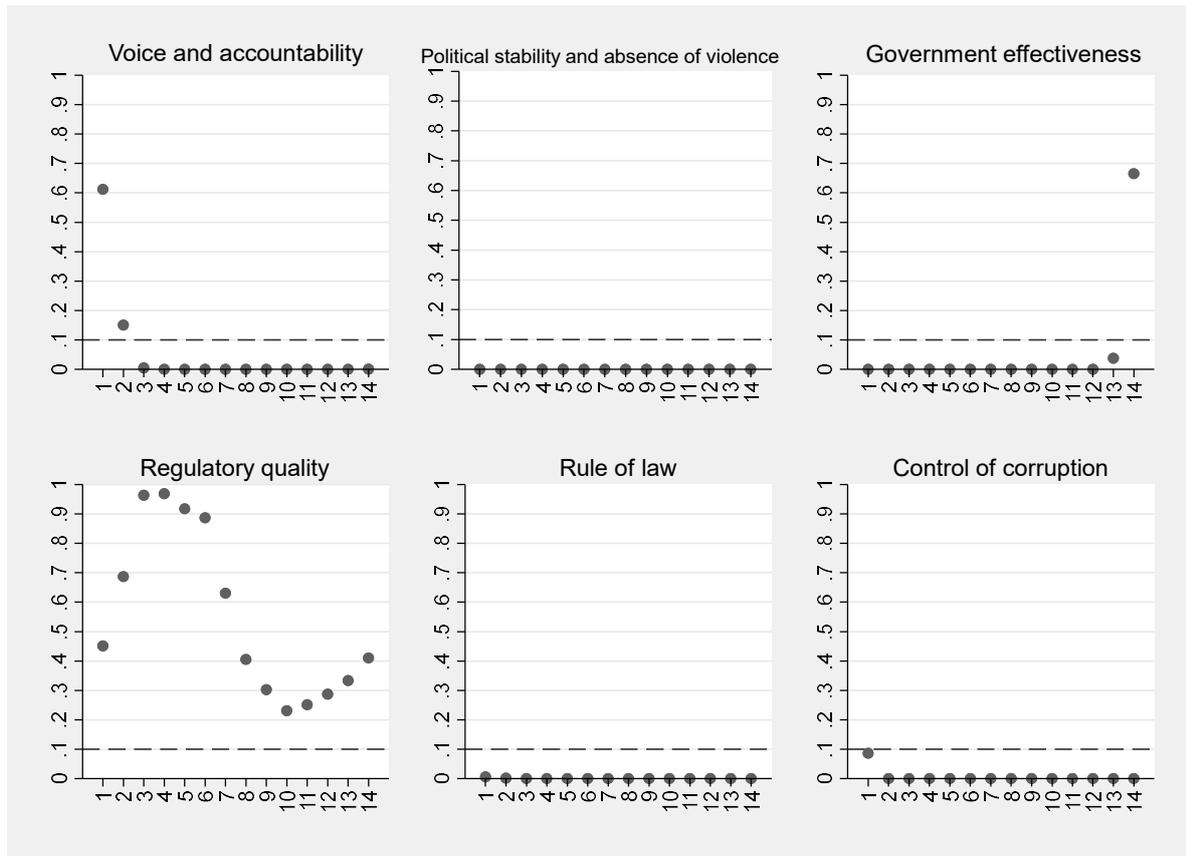

Figure 7: Inference on the institutional quality effect of staying out of the European Union, 2007-2020

Table 2: In-space placebo analysis of staying out of European Union

|  | Voice and accountability | Political stability and absence of violence | Government effectiveness | Regulatory quality | Rule of law | Control of corruption |
|---|---|---|---|---|---|---|
|  | (1) | (2) | (3) | (4) | (5) | (6) |
| $\lambda_{k=1,\text{full}}$ | -.061*** | -.401*** | -.135*** | -.085*** | -.145*** | -.200*** |
|  | (.012) | (.037) | (.013) | (.011) | (.017) | (.021) |
| 95% confidence bounds | {-.087, -.036} | {-.475, -.327} | {-.162, -.108} | {-.092, -.046} | {-.180, -.110} | {-.242, -.159} |
|  |  |  |  |  |  |  |
| # observations | 5,525 | 5,525 | 5,525 | 5,525 | 5,525 | 5,525 |
| # treated regions | 26 | 26 | 26 | 26 | 26 | 26 |
| # control regions | 195 | 195 | 195 | 195 | 195 | 195 |
| R2 | 0.07 | 0.34 | 0.19 | 0.17 | 0.13 |  |
| # placebo averages | >34 billion | >34 billion | >34 billion | >34 billion | >34 billion | >34 billion |
| Permutation method | Random sampling | Random sampling | Random sampling | Random sampling | Random sampling | Random sampling |
| Region-fixed effects (p-value) | YES (0.000) | YES (0.000) | YES (0.000) | YES (0.000) | YES (0.000) | YES (0.000) |
| Time-fixed effects (p-value) | YES (0.000) | YES (0.000) | YES (0.000) | YES (0.000) | YES (0.000) | YES (0.000) |

*Notes*: the table reports post-hypothetical membership coefficients on institutional quality gaps estimated by synthetic control method. Each specification includes the full set of region-fixed effects and time-fixed effects. Standard of the estimated placebo gap coefficients are adjusted for arbitrary heteroscedasticity and serially correlated stochastic disturbances using finite-sample adjustment of the empirical distribution function with error component model. Cluster-robust standard errors are denoted in the parentheses. Asterisks denote statistically significant coefficients at 10% (*), 5% (**), and 1% (***), respectively.

Table 3: Testing the equality of effect outcome distributions before and after the treatment

| | Voice and accountability | Political stability and absence of violence | Government effectiveness | Regulatory quality | Rule of law | Control of corruption |
|---|---|---|---|---|---|---|
| | (1) | (2) | (3) | (4) | (5) | (6) |
| $\Delta y$ | -.032 | -2.041 | -1.757 | -1.639 | -1.021 | -0.918 |
| t-stat | -0.37 | -1.60 | -5.123 | -2.886 | -1.620 | -1.229 |
| (p-value) | {0.711} | {0.121} | {0.000} | {0.008} | {0.118} | {0.711} |
| Kolmogorov-Smirnov equality of distributions test (p-value) | {0.761} | {0.039} | {0.002} | {0.016} | {0.111} | {0.761} |
| # obs | 26 | 26 | 26 | 26 | 26 | 26 |

Notes: the tables reports the equality of estimated gap distributions between the province under the formerly Habsburg rule and those outside of the Habsburg regime using two-sample Kolmogorov (1933) and Smirnov (1948) non-parametric test of the equality of continuous probability distributions. Glivenko-Cantelli p-values on the distance test statistics for the target empirical distribution function are obtained from the Jacobi theta function with the asymptotic distribution under the null hypothesis, and reported in the parentheses.

Figure 8: Transmission Mechanisms

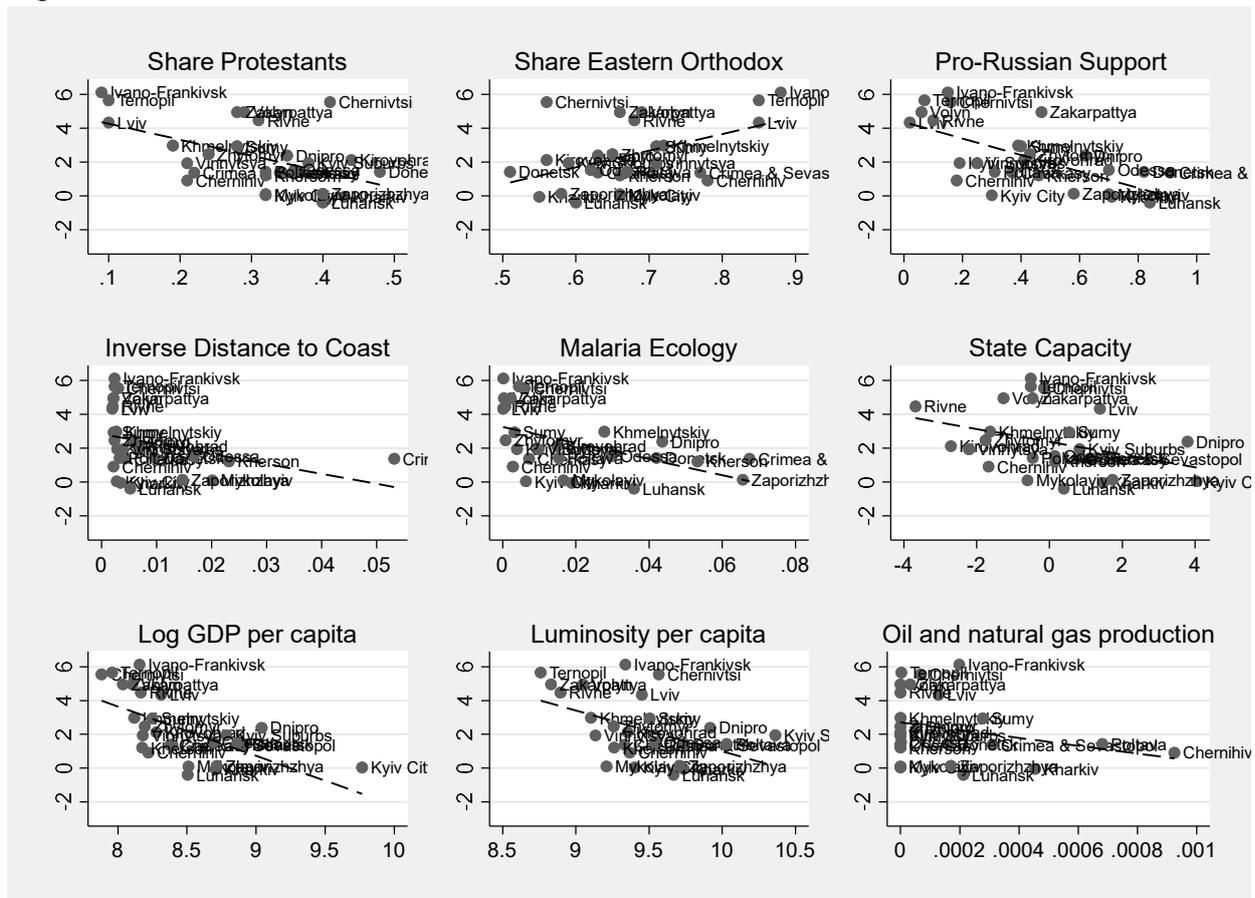